\documentclass[a4paper,11pt]{article}
\pdfoutput=1 

\usepackage{jheppub} 

\usepackage[T1]{fontenc} 
\usepackage{braket}

\makeatletter
\gdef\@fpheader{}
\makeatother

\preprint{OU-HET 1066}

\title{New scenario for aligned Higgs couplings originated from the twisted custodial symmetry at high energies}

\author{Masashi Aiko}
\author{and Shinya Kanemura}

\affiliation{Department of Physics, Osaka University, Toyonaka, Osaka 560-0043, Japan}

\emailAdd{m-aikou@het.phys.sci.osaka-u.ac.jp}
\emailAdd{kanemu@het.phys.sci.osaka-u.ac.jp}

\abstract{We investigate a new scenario of the two Higgs doublet model, where the current experimental data for the electroweak rho parameter and those for the Higgs boson couplings can be simultaneously explained.
In this scenario, the two Higgs doublet model is supposed to be a low energy effective theory up to a high energy scale $\Lambda$, above which a fundamental theory should appear.
It is assumed that the Higgs potential respects a global symmetry at $\Lambda$ (the twisted custodial symmetry), which is to be given as a consequence of the global symmetry structure of the fundamental theory above $\Lambda$.
By the analysis using one-loop renormalization group equations, the above experimental data can be explained in a natural way even when the masses of the extra Higgs bosons are near the electroweak scale.
We also discuss the predictions on the mass spectrum of the additional Higgs bosons and also those on the coupling constants of the standard-model-like Higgs boson, which make it possible to test this scenario at the current and future collider experiments.}

\begin{document} 
\maketitle
\flushbottom

\section{Introduction}
\label{sec:intro}
Since the discovery of the new particle $h(125)$ with the mass of $125$ GeV \cite{Aad:2012tfa, Chatrchyan:2012ufa}, it has turned out that its properties are in agreement with those of the Higgs boson in the standard model (SM) within uncertainties of the current data \cite{ATLAS:2019slw, Sirunyan:2018koj}.
Although no signal for new physics beyond the SM (BSM) has been observed at the LHC up to now, it is widely believed that the SM must be replaced by a more fundamental theory because of the following reasons.
First of all, the SM does not contain gravity.
Second, there is no unified description for the gauge groups and the flavor structure.
Third, the SM suffers from the hierarchy problem.
Finally, there are phenomena which cannot be explained within the SM, such as dark matter, baryon asymmetry of the universe and tiny neutrino masses.
 
While the Higgs boson $h(125)$ was found, the structure of the Higgs sector remains unknown.
There is no theoretical principle to insist on the minimal structure of the Higgs sector as introduced in the SM.
Actually, non-minimal Higgs sectors are often introduced in various new physics models, where above mentioned problems are tried to be solved.
These non-minimal Higgs sectors are characterized by the following properties;
1) the number of Higgs fields and their representations under the SM gauge symmetry,
2) the global symmetry structure of the Higgs potential, and
3) typical mass scales of additional Higgs bosons.
The current and future experiments are expected to reveal the structure of the Higgs sector through precision measurements of the discovered $h(125)$ particle and also direct searches of new particles.
By reconstructing the Higgs sector experimentally, the direction of new physics can be determined.

For a long time, several important experimental constraints have been known on the extended Higgs sector,  such as the suppression of flavor changing neutral currents (FCNCs) and the electroweak rho parameter is close to unity \cite{Tanabashi:2018oca}.
After the Higgs boson discovery, it has also turned out that the couplings of $h(125)$ with various SM particles are consistent with the predictions in the SM under the current experimental and theoretical uncertainties \cite{ATLAS:2019slw, Sirunyan:2018koj}.
This {\it alignment}, together with other data such as from LEP \cite{Schael:2006cr, Schael:2010aw, Abbiendi:2013hk}, LHC \cite{Aaboud:2017sjh, Aaboud:2018cwk, Aad:2019zwb} and flavor experiments \cite{Arbey:2017gmh, Misiak:2017bgg, Haller:2018nnx}, severely constrains the nature of the Higgs sector.

It is an important question how we can explain the above experimental constraints in the extended Higgs models.
One simple solution is to consider the {\it decoupling} of additional degree of freedoms.
In this scenario, the typical mass scale of new particles is so high that effects of the new physics on the electroweak observables are suppressed by the decoupling theorem \cite{Appelquist:1974tg}.
As an alternative, we may consider another scenario where some mechanism predicts  extended Higgs models which are phenomenologically SM like without decoupling.
This non-decoupling scenario would be motivated, for example, to realize the strong first order phase transition \cite{Grojean:2004xa, Kanemura:2004mg} which are required for successful electroweak baryogenesis \cite{Kuzmin:1985mm}, to explain the relic abundance of dark matter by weak interacting massive scalars \cite{Silveira:1985rk, McDonald:1993ex, Deshpande:1977rw, Barbieri:2006dq}, and to radiatively generate the neutrino tiny masses \cite{Ma:2006km, Aoki:2008av, Gustafsson:2012vj}.
Furthermore, this scenario can be tested at the current LHC experiments and future collider experiments such as the high-luminocity upgrade of the LHC (HL-LHC) \cite{Cepeda:2019klc} and the International Linear Collider (ILC) experiments \cite{Fujii:2015jha, Fujii:2017vwa}.

In this paper, we consider the two Higgs doublet model (2HDM) \cite{Lee:1973iz, Gunion:1989we, Branco:2011iw} as the concrete example.
This model is one of the well-motivated extensions of the SM, and provides interesting phenomenology such as the CP-violation in the scalar sector, strong first order phase transition, stable new scalar as a candidate of dark matter and so on.
Although the general 2HDM also suffers from above mentioned problems, it is widely known that one can naturally explain the current data for low energy observables if the Higgs sector respects some symmetries such as the $\mathbb{Z}_{2}$ symmetry \cite{Glashow:1976nt, Paschos:1976ay} or the custodial symmetry \cite{Sikivie:1980hm, Pomarol:1993mu, Gerard:2007kn, deVisscher:2009zb}.
The global symmetry structure of the Higgs potential in the 2HDM has been studied in Refs.~\cite{Deshpande:1977rw, Ginzburg:2004vp, Ferreira:2009wh, Battye:2011jj, Pilaftsis:2011ed}.
Among possible symmetries, we focus on the twisted custodial symmetry which was presented in Ref.~\cite{Gerard:2007kn}.
If the Higgs potential respects this global symmetry at the electroweak scale, one can explain $\rho\simeq 1$ and the $h(125)$ couplings to be SM like at the same time, where $\rho$ represents electroweak rho parameter.
However, it is rather unlikely that the Higgs potential exactly respects the twisted custodial symmetry at the electroweak scale, because this global symmetry is not the symmetry of the whole theory, and it is broken under the renormalization group (RG) evolution.

We here assume that the 2HDM is the low energy effective theory up to a high energy scale $\Lambda$, above which a more fundamental theory should appear.
In this scenario, the structure of the effective 2HDM shall reflect the nature of the theory at higher energy.
This type of the scenarios have been discussed in the different contexts in Refs.~\cite{Dev:2014yca, Gori:2017qwg}.
The realizations of the alignment from the concrete higher-energy models are discussed in Refs.~\cite{Coyle:2019exn, Benakli:2018vqz, Benakli:2018vjk}.
We make the assumption that the Higgs potential respects the twisted custodial symmetry at the high scale $\Lambda$ without a specific high energy theory in mind.
In this scenario, we examine whether $\rho \simeq 1$ and approximately aligned Higgs boson couplings can be naturally realized at the electroweak scale without decoupling of additional Higgs bosons.
This scenario gives the following distinctive predictions at the electroweak scale.
The CP-odd Higgs boson tends to be heavier than other Higgs bosons.
Masses of the additional CP-even Higgs boson and the charged Higgs bosons are almost degenerate.
Furthermore, if $\Lambda$ is as large as the Planck scale, the difference of squared masses between the CP-odd Higgs boson and the second CP-even Higgs boson converges to a definite value.
We also find that the several percent of modifications of the couplings between the SM-like Higgs boson and fermons are predicted if $\Lambda$ is close to the Planck scale, while the couplings between the SM-like Higgs boson and gauge bosons take SM-like values.
Therefore, this scenario can be tested through the direct search of additional Higgs bosons and the precision measurement of the SM-like Higgs boson couplings at the current and future collider experiments.

This paper is organized as follows.
In section~\ref{sec:model}, we review 2HDMs and the twisted custodial symmetry proposed in Ref.~\cite{Gerard:2007kn}.
In section~\ref{sec:conditions}, we describe our strategy of the numerical analysis and present violations of the twisted custodial symmetry generated under the RG evolution from the scale $\Lambda$ to the electroweak scale.
In section~\ref{sec:predictions}, we show predictions for the mass spectrum of additional Higgs bosons and the SM-like Higgs boson couplings.
The discussion and conclusions are given in section~\ref{sec:summary}.
We summarize the useful formulae of the parameters of the Higgs potential in appendix \ref{Apendix: Higgs-basis-params}. The one-loop RG equations of 2HDMs are summarized in appendix \ref{Apendix: RGE}.

\section{Two Higgs doublet models and the twisted custodial symmetry}
\label{sec:model}
In the 2HDM, we have two $SU(2)_{L}$ doublet fields $\Phi_{1}$ and $\Phi_{2}$ with a hypercharge $Y=1/2$.
In the most general 2HDM, FCNCs can appear at tree level and it is severely constrained by the experimental data.
If the Higgs sector respects the (softly-broken) discrete $\mathbb{Z}_{2}$ symmetry; $\Phi_{1}\rightarrow \Phi_{1}, \Phi_{2}\rightarrow -\Phi_{2}$, the tree-level FCNCs are prohibited \cite{Glashow:1976nt, Paschos:1976ay}.
According to the $\mathbb{Z}_{2}$ charge assignment of right-handed fermions given in Table~\ref{tab: Z2}, the 2HDM can be classified into four independent models, {\it i.e.} Type-I, Type-II, Type-X and Type-Y \cite{Barger:1989fj, Aoki:2009ha}.
In this paper, we consider the softly-broken $\mathbb{Z}_{2}$ symmetric Higgs sector%
\footnote{There is another approach where the flavor alignment ansatz is assumed, and two sets of the Yukawa matrices are proportional so that the tree-level FCNCs are eliminated \cite{Pich:2009sp}.
The stability of the flavor alignment ansatz under the RG evolution was studied in Ref. \cite{Gori:2017qwg}.}. 

\begin{table}[t]
\centering
  \begin{tabular}{|l|ccccccc||ccc|} \hline
  \multicolumn{1}{|l|}{ } & \multicolumn{7}{c||}{$\mathbb{Z}_{2}$ charge} & \multicolumn{3}{c|}{Mixing factor}\\ \cline{2-11}
  & $\Phi_{1}$ & $\Phi_{2}$ & $Q_{L}$ & $L_{L}$ & $u_{R}$ & $d_{R}$ & $e_{R}$ & $\xi_{u}$ & $\xi_{d}$ & $\xi_{e}$\\ \hline
    Type-I & $+$ & $-$ & $+$ & $+$ & $-$ & $-$ & $-$ & $\cot{\beta}$ & $\cot{\beta}$ & $\cot{\beta}$ \\
    Type-II & $+$ & $-$ & $+$ & $+$ & $-$ & $+$ & $+$ & $\cot{\beta}$ & $-\tan{\beta}$ & $-\tan{\beta}$ \\
    Type-X & $+$ & $-$ & $+$ & $+$ & $-$ & $-$ & $+$ & $\cot{\beta}$ & $\cot{\beta}$ & $-\tan{\beta}$ \\ 
    Type-Y & $+$ & $-$ & $+$ & $+$ & $-$ & $+$ & $-$ & $\cot{\beta}$ & $-\tan{\beta}$ & $\cot{\beta}$\\ \hline
    \end{tabular}
   \caption{Charge assignment of the softly-broken $\mathbb{Z}_{2}$ symmetry and the mixing factors in Yukawa interactions.}
   \label{tab: Z2}
\end{table}

\subsection{2HDMs with the softly-broken $\mathbb{Z}_{2}$ symmetry}
In the softly-broken $\mathbb{Z}_{2}$ symmetric scenario, the Higgs potential is given by
\begin{align}
V(\Phi_{1}, \Phi_{2})
&=
m_{11}^{2}\Phi_{1}^{\dagger}\Phi_{1} + m_{22}^{2}\Phi_{2}^{\dagger}\Phi_{2}-(m_{12}^{2}\Phi_{1}^{\dagger}\Phi_{2}+h.c.) \notag \\
&\quad
+\frac{1}{2}\lambda_{1}(\Phi_{1}^{\dagger}\Phi_{1})^{2}+\frac{1}{2}\lambda_{2}(\Phi_{2}^{\dagger}\Phi_{2})^{2}
+\lambda_{3}(\Phi_{1}^{\dagger}\Phi_{1})(\Phi_{2}^{\dagger}\Phi_{2})
+\lambda_{4}(\Phi_{1}^{\dagger}\Phi_{2})(\Phi_{2}^{\dagger}\Phi_{1}) \notag \\
&\quad
+\frac{1}{2}\left[\lambda_{5}(\Phi_{1}^{\dagger}\Phi_{2})^{2}+h.c.\right], \label{pot_thdm}
\end{align}
where $m_{11}^{2}, m_{22}^{2}$ and $\lambda_{1-4}$ are real parameters while $m_{12}^{2}$ and $\lambda_{5}$ are complex in general.
In the following, we analyze the CP conserving Higgs sector. This additional assumption is required when we consider the custodial symmetric Higgs sector as we discuss later.

It is useful to work in the Higgs basis \cite{Davidson:2005cw} to study the SM-like limit in 
the 2HDM;
\begin{align}
\left(\begin{array}{c}
H_{1} \\
H_{2}
\end{array}\right)
=
\left(\begin{array}{cc}
\cos{\beta} & \sin{\beta} \\
-\sin{\beta} & \cos{\beta}
\end{array}\right)
\left(\begin{array}{c}
\Phi_{1} \\
\Phi_{2}
\end{array}\right),
\end{align}
where the mixing angle is defined by $\tan{\beta}=v_{2}/v_{1}\ (0\le \beta\le \pi/2)$, and $v_{i}\ (i=1,2)$ are the vacuum expectation values (VEVs) of the neutral components of doublets in the $\mathbb{Z}_{2}$ basis given in Eq.~\eqref{pot_thdm}; $\braket{\Phi_{i}^{0}}=v_{i}/\sqrt{2}$.
In the Higgs basis, only one of the Higgs doublets, $H_{1}$, has the VEV, $v=\sqrt{v_{1}^{2}+v_{2}^{2}}=(\sqrt{2}G_{F})^{-1/2}\simeq 246$ GeV where $G_{F}$ is the Fermi constant.
We parameterize the doublets by
\begin{align}
H_{1}=\left(\begin{array}{c}
G^{+} \\
\frac{1}{\sqrt{2}}(v+h_{1}+iG)
\end{array}\right), \quad
H_{2}=\left(\begin{array}{c}
H^{+} \\
\frac{1}{\sqrt{2}}(h_{2}+iA)
\end{array}\right).
\end{align}

In the Higgs basis, the Higgs potential can be expressed as
\begin{align}
V(H_{1}, H_{2})
&=
 Y_{1}^{2}H_{1}^{\dagger}H_{1}+Y_{2}^{2}H_{2}^{\dagger}H_{2} - Y_{3}^{2}(H_{1}^{\dagger}H_{2}+H_{2}^{\dagger}H_{1}) \notag \\
&\quad
+\frac{1}{2}Z_{1}(H_{1}^{\dagger}H_{1})^{2}+\frac{1}{2}Z_{2}(H_{2}^{\dagger}H_{2})^{2}
+Z_{3}(H_{1}^{\dagger}H_{1})(H_{2}^{\dagger}H_{2})+Z_{4}(H_{1}^{\dagger}H_{2})(H_{2}^{\dagger}H_{1}) \notag \\
&\quad
+\left\{\frac{1}{2}Z_{5}(H_{1}^{\dagger}H_{2})^{2}+
\left[Z_{6}H_{1}^{\dagger}H_{1} + Z_{7}H_{2}^{\dagger}H_{2}\right]H_{1}^{\dagger}H_{2} + h.c.\right\},
\end{align}
where $Y_{i}^{2}$ and $Z_{i}$ are functions of $m_{ij}^{2}$ and $\lambda_{i}$.
The explicit formulae of $Y_{i}^{2}$ and $Z_{i}$ in terms of the parameters in the $\mathbb{Z}_{2}$ basis are given in appendix~\ref{Apendix: Higgs-basis-params}. The stationary conditions are given by
\begin{align}
Y_{1}^{2} = -\frac{1}{2}Z_{1}v^{2}, \quad Y_{3}^{2} = \frac{1}{2}Z_{6}v^{2}.
\end{align}
We note that not all of parameters in the Higgs basis are independent in the softly-broken $\mathbb{Z}_{2}$ symmetric scenario as discussed in appendix~\ref{Apendix: Higgs-basis-params}.

The mass matrices of the charged states and the CP-odd states are diagonalized in the Higgs basis,
\begin{align}
m_{H^{\pm}}^{2} &= Y_{2}^{2}+\frac{1}{2}Z_{3}v^{2}=M^{2}-\frac{1}{2}(\lambda_{4}+\lambda_{5})v^{2}, \\
m_{A}^{2} &= Y_{2}^{2}+\frac{1}{2}(Z_{3}+Z_{4}-Z_{5})v^{2}=M^{2}-\lambda_{5}v^{2},
\end{align}
where we have introduced the softly $\mathbb{Z}_{2}$ breaking scale $M=m_{12}/\sqrt{\sin{\beta}\cos{\beta}}$. The mass matrix for the CP-even states is not diagonarazed in the Higgs basis,
\begin{align}
\frac{1}{2}
\left(\begin{array}{cc}
h_{1},\ h_{2}
\end{array}\right)
{\cal M}^{2}
\left(\begin{array}{c}
h_{1} \\
h_{2}
\end{array}\right)
&=
\frac{1}{2}
\left(\begin{array}{cc}
h_{1},\ h_{2}
\end{array}\right)
\left(\begin{array}{cc}
Z_{1}v^{2} & Z_{6}v^{2} \\
Z_{6}v^{2} & Y_{2}^{2}+\frac{1}{2}Z_{345}v^{2}
\end{array}\right)
\left(\begin{array}{c}
h_{1} \\
h_{2}
\end{array}\right),
\end{align}
and we need further rotation to obtain CP-even mass eigenstates $h$ and $H$,
\begin{align}
\left(\begin{array}{c}
H \\
h
\end{array}\right)
=
\left(\begin{array}{cc}
\cos{(\beta-\alpha)} & -\sin{(\beta-\alpha)} \\
\sin{(\beta-\alpha)} & \cos{(\beta-\alpha)}
\end{array}\right)
\left(\begin{array}{c}
h_{1} \\
h_{2}
\end{array}\right).
\end{align}
The squared masses of the CP-even Higgs bosons and the mixing angle $\beta-\alpha$ are given by
\begin{align}
m_{H}^{2} &= \cos^{2}{(\beta-\alpha)}{\cal M}_{11}^{2}+\sin^{2}{(\beta-\alpha)}{\cal M}_{22}^{2}-\sin{2(\beta-\alpha)}{\cal M}_{12}^{2}, \\
m_{h}^{2} &= \sin^{2}{(\beta-\alpha)}{\cal M}_{11}^{2}+\cos^{2}{(\beta-\alpha)}{\cal M}_{22}^{2}+\sin{2(\beta-\alpha)}{\cal M}_{12}^{2}, \\
\tan&{2(\beta-\alpha)} = \frac{-2{\cal M}_{12}^{2}}{{\cal M}_{11}^{2}-{\cal M}_{22}^{2}}.
\end{align}
We use the convention where $\sin{(\beta-\alpha)}$ is always positive, {\it i.e.} $0\le \beta-\alpha \le \pi$, and $\cos{(\beta-\alpha)}$ has the opposite sign from $Z_{6}$ \cite{Bernon:2015qea}.
In this paper, we identify $h$ as the discovered Higgs boson $h(125)$, and all additional scalar bosons are assumed to be heavier than $h(125)$.

\subsection{Alignment limit}
In the mass eigenstate, the interaction terms among the gauge bosons and the CP-even scalars are given by,
\begin{align}
{\cal L}_{int}
=
[\sin{(\beta-\alpha)}h+\cos{(\beta-\alpha)}H]\left(\frac{m_{W}^{2}}{v}W^{+\mu}W_{\mu}^{-}+\frac{m_{Z}^{2}}{2v}Z^{\mu}Z_{\mu}\right). \label{int_gauge}
\end{align}
The Yukawa interaction terms among the fermions and the CP-even scalars are given by \begin{align}
{\cal L}_{int}
=
-\sum_{f=u,d,e}\frac{m_{f}}{v}\left(\xi_{h}^{f}\overline{f}fh+\xi_{H}^{f}\overline{f}fH\right),
\label{int_fermion}
\end{align}
where
\begin{align}
\xi_{h}^{f}&=\sin{(\beta-\alpha)}+\xi_{f}\cos{(\beta-\alpha)}, \\
\xi_{H}^{f}&=\cos{(\beta-\alpha)}-\xi_{f}\sin{(\beta-\alpha)},
\end{align}
and $\xi_{f}$ is the type-dependent parameter given in Table \ref{tab: Z2}.
When $\sin{(\beta-\alpha)}=1$, the couplings of $h$ with various SM particles become SM like. We call this SM-like limit, $\sin{(\beta-\alpha)}=1$, as the alignment limit in this paper.

The alignment limit can be achieved in the different two ways \cite{Kanemura:1997wx, Gunion:2002zf, Kanemura:2004mg, Carena:2013ooa}: (i) decoupling of additional Higgs bosons, and (ii) alignment without decoupling.

In the scenario (i), we take the decoupling limit: ${\cal M}_{22}^{2}\simeq M^{2} \gg {\cal M}_{11}^{2}, {\cal M}_{12}^{2}$. Then, we have
\begin{align}
\cos{2(\beta-\alpha)}
=
\frac{{\cal M}_{11}^{2}-{\cal M}_{22}^{2}}{\sqrt{({\cal M}_{11}^{2}-{\cal M}_{22}^{2})^{2}+(-2{\cal M}_{12}^{2})^{2}}}
\simeq
-1. \label{decouple}
\end{align}
Eq.~\eqref{decouple} indicates $\beta-\alpha\simeq \pi/2$, and the couplings of $h$ become SM like. In this scenario, the masses of the additional Higgs bosons are close to $M$, and they are decoupled from the electroweak physics.

In the scenario (ii), off-diagonal component of the mass matrix for the CP-even states is equal to zero;
\begin{align}
Z_{6}=-\frac{1}{2}\sin{2\beta}\big[\lambda_{1}\cos^{2}{\beta}-\lambda_{2}\sin^{2}{\beta}-\lambda_{345}\cos{2\beta}\big]=0, \label{align_wo_dec}
\end{align}
where we have used the abbreviation $\lambda_{345}=\lambda_{3}+\lambda_{4}+\lambda_{5}$.
In this scenario, the additional Higgs bosons need not to be decoupled, and the masses of these particles can be taken around the electroweak scale. Therefore, this scenario is testable in the current and future experiments \cite{Kanemura:2014bqa, Kanemura:2014dea}.
The simple realization of the condition in Eq.~\eqref{align_wo_dec} is taking the natural alignment conditions \cite{Dev:2014yca},
\begin{align}
\lambda_{1}=\lambda_{2}=\lambda_{345}, \label{natu_align}
\end{align}
where the alignment is realized independently of the value of $\tan{\beta}$.
We will see that the Higgs quartic couplings satisfy the natural alignment conditions given in Eq.~\eqref{natu_align} if the Higgs potential respects the twisted custodial symmetry \cite{Gerard:2007kn} with $\beta \neq 0, \pi/4$ or $\pi/2$.

\subsection{Oblique parameters}
The effect of new physics on the electroweak precision observables can be parameterized in terms of the oblique parameters, $S, T$ and $U$ \cite{Peskin:1990zt, Peskin:1991sw}. In the 2HDM, the oblique parameters are modified from those in the SM due to the additional Higgs bosons loop contritions and modified SM-like Higgs boson couplings.

Among these oblique parameters, the $T$ parameter is related to the rho parameter as
$\rho=1+\alpha_{em}T$, and it is sensitive for the mass squared differences of the Higgs bosons. When we decompose the $T$ parameter into the SM contribution $T_{{\rm SM}}$ and the new physics effects $\Delta T$, $\Delta T$ is given by \cite{Toussaint:1978zm, Bertolini:1985ia, Grimus:2008nb, Kanemura:2011sj}
\begin{align}
\Delta T &= \frac{1}{16\pi^{2}\alpha_{{\rm em}}v^{2}}\bigg\{F(m_{H^{\pm}}^{2}, m_{A}^{2})
+s_{\beta-\alpha}^{2}\Big[F(m_{H^{\pm}}^{2}, m_{H}^{2})-F(m_{H}^{2}, m_{A}^{2})\Big] \notag \\
&+c_{\beta-\alpha}^{2}\Big[F(m_{h}^{2}, m_{H^{\pm}}^{2})-F(m_{h}^{2}, m_{A}^{2})\Big]
+3c_{\beta-\alpha}^{2}\Big[F(m_{H}^{2}, m_{Z}^{2})-F(m_{H}^{2}, m_{W}^{2})\Big] \notag \\
&+3s_{\beta-\alpha}^{2}\Big[F(m_{h}^{2}, m_{Z}^{2})-F(m_{h}^{2}, m_{W}^{2})\Big]
-3\Big[F(m_{h_{ref}}^{2}, m_{Z}^{2})-F(m_{h_{ref}}^{2}, m_{W}^{2})\Big]
\bigg\},
\end{align}
where we have used abbreviations of $s_{\beta-\alpha}=\sin{(\beta-\alpha)}$ and $c_{\beta-\alpha}=\cos{(\beta-\alpha)}$. The function $F(x,y)$ is defined by
\begin{align}
F(x,y) = \frac{x+y}{2}-\frac{xy}{x-y}\ln{\frac{x}{y}},
\end{align}
and $F(x,x) = 0$. If one of the following relations; (A) , (B) or (C) is satisfied, the loop corrections due to the additional Higgs bosons are canceled, and $\Delta T$ becomes small;
\begin{align}
({\rm A}):&\quad m_{A}=m_{H^{\pm}}, \\
({\rm B}):&\quad m_{H}=m_{H^{\pm}}\  \text{and}\  \sin{(\beta-\alpha)}=1, \label{condi_B}\\
({\rm C}):&\quad m_{h}=m_{H^{\pm}}\  \text{and}\  \cos{(\beta-\alpha)}=1.
\end{align}
The possible value of $\Delta T$ is strictly constrained by the electroweak precision data \cite{Tanabashi:2018oca, Haller:2018nnx}, and we expect that one of the above conditions is realized at the electroweak scale at least approximately.
We will see that the condition (B) is derived as a consequence of the twisted custodial symmetry in the Higgs potential, and we can understand the smallness of $\Delta T$ in terms of the global symmetry structure of the Higgs potential.\footnote{We note that opposite statement is not true.
The condition (B) does not imply the presence of the twisted custodial symmetry in the Higgs potential.
As we discuss in Sec.~\ref{subsec: tcs}, we need further degeneracy among the masses of charged Higgs boson, the additional CP-even Higgs boson and the softly $Z_{2}$ breaking scale $M$ to realize the twisted custodial symmetry.}

\subsection{Twisted custodial symmetry in the 2HDMs}
\label{subsec: tcs}
We introduce bi-doublet fields \cite{Sikivie:1980hm, Pomarol:1993mu, Gerard:2007kn, Haber:2010bw} to study the structure of the Higgs potential especially for the $SU(2)_{L}\times SU(2)_{R}$ global symmetry.
\begin{align}
M_{i} = (i\sigma_{2}H_{i}^{*}, H_{i}), \quad (i=1,2),
\end{align}
where $\sigma_{2}$ is the second matrix of the Pauli matrices $\sigma_{a}\ (a=1,2,3)$.
These bi-doublet fields transform under the local gauge transformations as follows,
\begin{align}
SU(2)_{L}:\ M_{i} \rightarrow \exp{[ig\alpha_{a}(x)\tau_{a}]}M_{i}, \quad 
U(1)_{Y}:\ M_{i}\rightarrow M_{i}\exp{[-ig'Y\alpha_{4}(x)\sigma_{3}]},
\end{align}
where $\tau_{a}=\sigma_{a}/2$. We note that we may also use the following bi-doublets,
\begin{align}
M_{i}P \equiv M_{i}\exp{[-i\chi\sigma_{3}]}=M_{i}{\rm diag}(e^{-i\chi}, e^{i\chi}),\quad {\rm with}\ 0\le \chi < 2\pi,
\end{align}
to construct the gauge-invariant Higgs potential since the $U(1)_{Y}$ transformation of bi-doublet fields commutes with $P$.

In the Higgs basis, we defined $H_{1}$ such that the VEV of this field is real and positive.
Therefore, we consider $M_{1}$ and $M'_{2}=M_{2}P$ as building blocks of the Higgs potential. The $SU(2)_{L}\times SU(2)_{R}$ transformations of $M_{1}$ and $M'_{2}$ are given by
\begin{align}
M_{1} \rightarrow L M_{1}R^{\dagger}, \quad M'_{2} \rightarrow L M'_{2}R^{\dagger},
\end{align}
where $L\in SU(2)_{L}, R \in SU(2)_{R}$. We can construct four $SU(2)_{L}\times U(1)_{Y}$ invariants as
\begin{align}
{\rm Tr}(M_{1}^{\dagger}M_{1}) &= 2|H_{1}|^{2}, \\
{\rm Tr}({M'}_{2}^{\dagger}M'_{2}) &= 2|H_{2}|^{2}, \\
{\rm Tr}(M_{1}^{\dagger}M'_{2}) &= e^{i\chi}H_{1}^{\dagger}H_{2}+e^{-i\chi}H_{2}^{\dagger}H_{1}, \\
{\rm Tr}(M_{1}^{\dagger}M'_{2}\sigma_{3}) &= e^{i\chi}H_{1}^{\dagger}H_{2}-e^{-i\chi}H_{2}^{\dagger}H_{1},
\end{align}
where ${\rm Tr}(M_{1}^{\dagger}M_{1}), {\rm Tr}({M'}_{2}^{\dagger}M'_{2})$ and ${\rm Tr}(M_{1}^{\dagger}M'_{2})$ are hermitian and $SU(2)_{L}\times SU(2)_{R}$ invariant. On the other hand, ${\rm Tr}(M_{1}^{\dagger}M'_{2}\sigma_{3})$ is anti-hermitian and does not respect $SU(2)_{L}\times SU(2)_{R}$ symmetry.

The Higgs potential can be rewritten in terms of these invariants as
\begin{align}
V(M_{1}, M'_{2}) &= \frac{1}{2}Y_{1}^{2}{\rm Tr}(M_{1}^{\dagger}M_{1})+\frac{1}{2}Y_{2}^{2}{\rm Tr}({M'}_{2}^{\dagger}M'_{2})
-{\rm Re}(Y_{3}^{2}e^{-i\chi}){\rm Tr}(M_{1}^{\dagger}M'_{2}) \notag \\
&\quad
+\frac{1}{8}Z_{1}\,{\rm Tr}^2(M_{1}^{\dagger}M_{1})
+\frac{1}{8}Z_2\,{\rm Tr}^2({M'}_{2}^{\dagger}M'_{2})
+\frac{1}{4}Z_{3}\,{\rm Tr}(M_{1}^{\dagger}M_{1}^{}){\rm Tr}({M'}_{2}^{\dagger}M'_{2}) \notag \\
&\quad
+\frac{1}{4}[Z_4+{\rm Re}(Z_5e^{-2i\chi})]{\rm Tr}^2(M_{1}^{\dagger}M'_{2}) \notag\\
&\quad
+\frac{1}{2}[{\rm Re}(Z_6e^{-i\chi})\,{\rm Tr}(M_{1}^{\dagger}M_{1}^{})
+{\rm Re}(Z_7e^{-i\chi})\,{\rm Tr}({M'}_{2}^{\dagger}M'_{2})]{\rm Tr}(M_{1}^{\dagger}M'_{2}) 
\notag\\
&\quad 
-i\,{\rm Im}(Y^2_{3}e^{-i\chi})\,{\rm Tr}(M_{1}^{\dagger}M'_{2}\sigma_3) 
-\frac{1}{4}[Z_4-{\rm Re}(Z_5e^{-2i\chi})]
{\rm Tr}^2(M_{1}^{\dagger}M'_{2}\sigma_3) \notag\\  
&\quad
+\frac{i}{2}{\rm Im}(Z_5e^{-2i\chi})\,
{\rm Tr}(M_{1}^{\dagger}M'_{2}){\rm Tr}(M_{1}^{\dagger}M'_{2}\sigma_3) \notag \\    
&\quad
+\frac{i}{2}[{\rm Im}(Z_6e^{-i\chi})\,{\rm Tr}(M_{1}^{\dagger}M_{1}^{})
+{\rm Im}(Z_7e^{-i\chi})\,{\rm Tr}({M'}_{2}^{\dagger}M'_{2})]
{\rm Tr}(M_{1}^{\dagger}M'_{2}\sigma_3). \label{pot1}
\end{align}
As we mentioned, $Z_{6}$ and $Z_{7}$ are expressed in terms of $Z_{1-5}$ and $\tan{\beta}$ in the softly-broken $\mathbb{Z}_{2}$ symmetric scenario.

If we assume the global $SU(2)_L\times SU(2)_R$ symmetry, we obtain 
\begin{align}
{\rm Im}(Y^2_{3}e^{-i\chi})&={\rm Im}(Z_5e^{-2i\chi})={\rm Im}(Z_6e^{-i\chi})={\rm Im}(Z_7e^{-i\chi})=0, \label{cust-con1} \\
Z_4&={\rm Re}(Z_5e^{-2i\chi}). \label{cust-con2}
\end{align}
In order to satisfy Eqs.~\eqref{cust-con1} and \eqref{cust-con2}, CP invariance is required in the Higgs potential \cite{Haber:2010bw}.
When the Higgs potential is CP invariant, one can rephase the Higgs basis field $H_{2}$ so that the $Y_{i}^{2}$ and $Z_{i}$ are real, and we have
\begin{align}
Z_4&=Z_5   && {\rm for}\quad \chi=0,\,\pi, \label{cA} \\
&\text{or} \notag \\
Z_4&=-Z_5\quad {\rm and}\quad Y_{3}^{2}=Z_6=Z_7=0 && {\rm for}\quad \chi=\pi/2,\,3\pi/2. \label{cB}
\end{align}
where we have used Eq.~\eqref{cust-con1} to derive the second conditions of Eq.~\eqref{cB}.
These conditions can be expressed in terms of the parameters in Eq.~\eqref{pot_thdm}, respectively,  
\begin{align}
\lambda_4&=\lambda_5,  && {\rm for}\quad \chi=0, \pi, \label{cAp} \\
&\text{or} \notag \\
\lambda_4&=-\lambda_5,\ \lambda_1= \lambda_2 = \lambda_{3} && {\rm for}\quad \chi=\pi/2,\,3\pi/2. \label{cBp}
\end{align}
The former case shown in Eq.~\eqref{cAp} corresponds to the usual realization of the custodial symmetry $(m_{A}=m_{H^{\pm}})$ introduced in Ref.~\cite{Pomarol:1993mu},
and the latter case in Eq.~\eqref{cBp} is so-called the twisted custodial symmetry \cite{Gerard:2007kn}.
We can see that the natural alignment conditions given in Eq.~\eqref{natu_align} are realized if the Higgs potential respects the twisted custodial symmetry.
We note that the conditions in Eq.~\eqref{cBp} are relaxed if $\beta=0, \pi/4$ or $\pi/2$ as we discuss in Appendix~A.

In the twisted-custodial symmetric scenario, the Higgs potential is given by
\begin{align}
V(H_{1}, H_{2})
&=
 Y_{1}^{2}H_{1}^{\dagger}H_{1}+Y_{2}^{2}H_{2}^{\dagger}H_{2}
+\frac{1}{2}Z_{S}(H_{1}^{\dagger}H_{1}+H_{2}^{\dagger}H_{2})^{2}
-Z_{AS}(H_{1}^{\dagger}H_{2}-H_{2}^{\dagger}H_{1})^{2},
\end{align}
where we have introduced $Z_{S}=Z_{1}=Z_{2}=Z_{3}$ and $Z_{AS}=Z_{4}=-Z_{5}$.
The masses of the physical Higgs bosons are expressed by
\begin{align}
m_{h}^{2} &= Z_{S}v^{2}, \\
m_{H^{\pm}}^{2}&=m_{H}^{2}= M^{2}= Y_{2}+\frac{1}{2}Z_{S}v^{2}, \label{mHp} \\
m_{A}^{2} &= M^{2}+Z_{AS}v^{2}. \label{mA}
\end{align}
We note that all scalars are simultaneously diagonalized in the Higgs basis, and we can identify $h=h_{1}$ and $H=h_{2}$.
This indicates that $\sin{(\beta-\alpha)}=1$ in the twisted-custodial symmetric scenario.
As we discussed in the previous subsection, $\Delta T$ becomes small when $A$ is degenerate with $H^{\pm}$ or when $H$ is degenerate with $H^{\pm}$ in the alignment limit.
These conditions are naturally realized when the Higgs potential respects the custodial symmetry.
We note that in order to realize the small $T$ parameter only, the degeneracy between $M^{2}$ and $m_{H^{\pm}}^{2}(=m_{H}^{2})$ is not always needed.
This indicates that $Z_{7}$ can take a non-zero value while keeping the smallness of the $T$ parameter at the one-loop level.
However, we have finite scalar-loop contributions to $\Delta T$ at two-loop level in such a scenario \cite{Hessenberger:2016atw}.
In this paper, we do not consider such a scenario and, we concentrate on explaining the values of low energy observables in terms of the global symmetry of the Higgs potential.

As it is pointed out in Ref.~\cite{Haber:2010bw}, the CP quantum numbers of $H$ and $A$ cannot be determined only from the Higgs potential when $Z_{6}=Z_{7}=0$.
If neutral Higgs-fermion interactions are CP conserving, as the case we are considering, we can determine such that $H$ is CP-even and $A$ is CP-odd.
In the twisted-custodial symmetric scenario defined in Eq.~\eqref{cB}, $H^{\pm}$ and the CP-even scalar $H$ are degenerate in mass and this scenario is different from Case II in Ref.~\cite{Pomarol:1993mu} where $H$ should be regarded as the CP-odd state.
Therefore, we can treat $\tan{\beta}$ as a free parameter differently from Case II where $\tan{\beta}=1$ is required.

As it is well known that Yukawa coupling constants and the $U(1)_{Y}$ gauge coupling $g'$ violate the custodial symmetry, so that this global symmetry is not the symmetry of the whole 2HDM Lagrangian.
Therefore, the relations among the Higgs quartic couplings given in Eq.~\eqref{cBp} are broken under the RG evolution.
Although we can explain the observed data of $\Delta T$ and aligned Higgs boson couplings by the twisted custodial symmetry in the Higgs potential, those violations indicate the peculiarity of the scenario where the Higgs potential exactly respects the twisted custodial symmetry at the electroweak scale.
In the following sections, we investigate the possibility of the approximate realization of the twisted custodial symmetry at the electroweak scale, starting from a twisted-custodial symmetric theory at some higher scale $\Lambda$.

\section{Boundary conditions and other setup for our scenario}
\label{sec:conditions}
In this section, we discuss constraints on $S$ and $T$ parameters and the Higgs boson couplings in the twisted-custodial symmetric scenario at a high energy scale $\Lambda$.
We use the one-loop RG equations in the following analysis. The list of the one-loop RG equations can be found in appendix~\ref{Apendix: RGE}.

\subsection{Boundary conditions at $\Lambda$}
There are works in which several authors investigated the validity of the 2HDM up to higher energy scale and bounds of the masses of Higgs bosons through the RG evolution of the Higgs quartic couplings \cite{Nie:1998yn, Kanemura:1999xf, Ferreira:2009jb}.
After the Higgs boson discovery, the possible cutoff scale was examined under the current experimental data \cite{Chakrabarty:2014aya, Das:2015mwa, Basler:2017nzu}.
In these works, the experimental constraints on the oblique parameters and the SM-like Higgs boson couplings are satisfied as the initial conditions of the RG evolution at the electroweak scale.
We can study the structure of the Higgs potential along this line by assuming that a global symmetry is exactly realized at the electroweak scale.
However, such a scenario is not plausible unless the global symmetry is a symmetry of the whole theory, because the Higgs potential no longer respects the global symmetry at any other scale. 

In this paper, we investigate the possible explanation for the observed data at and below the electroweak scale in terms of the global symmetry of the Higgs potential at some higher scale $\Lambda$ above which a fundamental theory should appear.
Below $\Lambda$, the twisted-custodial symmetric 2HDM appears as the low energy effective theory.
Following this scenario, we impose the condition of Eq.~\eqref{cBp} at the scale $\Lambda$,
\begin{align}
\quad \lambda_4(\Lambda)=-\lambda_5(\Lambda),\ \lambda_1(\Lambda)= \lambda_2(\Lambda) = \lambda_{3}(\Lambda). \label{bd_lam}
\end{align}
As we have already mentioned, the conditions in Eq.~\eqref{bd_lam} can be relaxed if $\beta=0, \pi/4$ or $\pi/2$.
However we adopt the conditions in Eq.~\eqref{bd_lam} as the boundary conditions at the scale $\Lambda$ for simplicity.
The twisted custodial symmetry  in the Higgs potential is broken under the RG evolution from $\Lambda$ to the electroweak scale due to the corrections of $g'$ and $y_{i}$, so that we expect that the conditions in Eq.~\eqref{bd_lam} are broken at the electroweak scale.

\subsection{Theoretical and experimental bounds}
We numerically generate the parameters under the boundary conditions in Eq.~\eqref{bd_lam}.
We also impose the following theoretical conditions at and below $\Lambda$.
First, we require (a) vacuum stability conditions \cite{Deshpande:1977rw, Klimenko:1984qx, Sher:1988mj, Kanemura:1999xf, Ivanov:2006yq} which ensure that the Higgs potential is bounded from below in any direction with a large scalar field value. These conditions are given by
\begin{align}
&\lambda_{1}(\mu)>0, \ \lambda_{2}(\mu)>0, \notag \\
&\sqrt{\lambda_{1}(\mu)\lambda_{2}(\mu)}+\lambda_{3}(\mu)
+{\rm min}[0, \lambda_{4}(\mu)+\lambda_{5}(\mu), \lambda_{4}(\mu)-\lambda_{5}(\mu)]>0.
\end{align}
Second, we require (b) perturbative unitarity bound \cite{Kanemura:1993hm, Akeroyd:2000wc, Ginzburg:2005dt, Kanemura:2015ska} which imposes that all the independent eigenvalues of the $T$ matrix, $a_{i, \pm}^{0}\ (i=1 \text{-} 6)$, for the $S$-wave amplitude of the elastic scatterings of 2-body boson states satisfy
\begin{align}
\left|a_{i, \pm}^{0}\right|\le 1,
\end{align}
where each of $a_{i, \pm}^{0}$ is given by
\begin{align}
a_{1, \pm}^{0}&=\frac{1}{32\pi}\left[3(\lambda_{1}(\mu)+\lambda_{2}(\mu))\pm \sqrt{9(\lambda_{1}(\mu)-\lambda_{2}(\mu))^{2}+4(2\lambda_{3}(\mu)+\lambda_{4}(\mu))^{2}}\right], \\
a_{2,\pm}^{0}&=\frac{1}{32\pi}\left[(\lambda_{1}(\mu)+\lambda_{2}(\mu))\pm \sqrt{(\lambda_{1}(\mu)-\lambda_{2}(\mu))^{2}+4\lambda_{4}^{2}(\mu)}\right], \\
a_{3,\pm}^{0}&=\frac{1}{32\pi}\left[(\lambda_{1}(\mu)+\lambda_{2}(\mu))\pm \sqrt{(\lambda_{1}(\mu)-\lambda_{2}(\mu))^{2}+4\lambda_{5}^{2}(\mu)}\right], \\
a_{4, \pm}^{0}&=\frac{1}{16\pi}(\lambda_{3}(\mu)+2\lambda_{4}(\mu)\pm3\lambda_{5}(\mu)), \\
a_{5, \pm}^{0}&=\frac{1}{16\pi}(\lambda_{3}(\mu)\pm\lambda_{4}(\mu)), \\
a_{6,\pm}^{0}&=\frac{1}{16\pi}(\lambda_{3}(\mu)\pm\lambda_{5}(\mu)).
\end{align}
Finally, we also require (c) absence of the Landau pole;
\begin{align}
\lambda_{i}(\mu)<8\pi, \quad y_{f}^{2}(\mu)<4\pi.
\end{align}
To evaluate the $\lambda_{i}(\mu), (m_{Z}\le \mu < \Lambda)$ from $\lambda_{i}(\Lambda)$, we need to know the values of the gauge couplings $g_{i}(\Lambda)$ and Yukawa couplings $y_{i}(\Lambda)$.
At the one-loop level, the beta functions of $g_{i}$ are independent of both $\lambda_{i}$ and $y_{i}$, and therefore $g_{i}(\Lambda)$ can be evaluated from the inputs $g_{i}(m_{Z})$.
Furthermore, the beta functions of $y_{i}$ are independent of $\lambda_{i}$, so that we can evaluate $y_{i}(\Lambda)$ from the inputs values of $m_{f}$ and $\tan{\beta}$ at the electroweak scale.
For given values of $\lambda_{i}(\Lambda)$, $y_{i}(\Lambda)$ and $g_{i}(\Lambda)$, we calculate $\lambda_{i}(\mu), y_{i}(\mu)$ and  $g_{i}(\mu)$ iteratively by using the RG equations and confirm that the conditions (a), (b) and (c) are satisfied at each step until the electroweak scale. For the RG analysis, we include the contributions of third generation fermions, {\it i.e.} $t, b$ and $\tau$, and safely neglect the contributions of other generations.

To translate $\lambda_{i}(m_{Z})$ into the masses of the Higgs bosons and the mixing angle $\beta-\alpha$, we need to know the value of $M$.
We scan $M$ between $[0, 1000]$ GeV, and check whether the parameters satisfy (d) the global minimum condition \cite{Barroso:2012mj, Barroso:2013awa};
\begin{align}
D = m_{12}^{2}(m_{11}^{2}-k^{2}m_{22}^{2})\left(\frac{v_{2}}{v_{1}}-k\right) > 0,
\end{align}
where $k=\sqrt[4]{\lambda_{1}/\lambda_{2}}$.
We extract parameters which reproduce the mass of the discovered Higgs boson $m_{h}\simeq 125$ GeV.
We allow the 5$\%$ error on the deduced value of $m_{h}$ instead of imposing a more strict constraint, because our numerical analysis is done at the one-loop order.

We evaluate $S$ and $T$ parameters and confirm whether the predicted values are consistent with the current experimental data; $S=0.02\pm 0.10, T=0.07 \pm 0.12$ and the $92\%$ correlation among them \cite{Tanabashi:2018oca}.
We require the agreement between the predictions and observed data to be at the $2\sigma$ level \cite{Giardino:2013bma}.

The mixing of the CP-even scalars, $\cos{(\beta-\alpha)}$, is evaluated from $\lambda_{i}(m_{Z}), M$ and $\tan{\beta}$. We check whether the predicted value of $\cos{(\beta-\alpha)}$ satisfies the current experimental bound at the $2\sigma$ level \cite{ATLAS:2019slw, Sirunyan:2018koj}.

\subsection{Violation of the twisted custodial symmetry at the electroweak scale}
In the twisted-custodial symmetric scenario, smallness of the $\Delta T$ parameter and the SM-like Higgs boson couplings are realized by the conditions in Eq.~\eqref{cBp}.
However, these conditions would be violated at the electroweak scale even if the Higgs potential respects the twisted custodial symmetry at the high scale $\Lambda$.
In this subsection, we analyze the violation of the conditions in Eq.~\eqref{cBp} under the current experimental data.

In the following discussion, we show the results in the case of $\tan{\beta}=5$ in Type-I as a representative.
We have checked that the results for Type-X and Type-Y are similar for the those for Type-I and Type-II, respectively. Furthermore, the difference between Type-I and Type-II mainly comes from the $b\rightarrow s\gamma$ constraint on the charged Higgs boson in Type-II: $m_{H^{\pm}}\gtrsim 580$ GeV \cite{Misiak:2017bgg}.
As we will see later, in our scenario, $H^{\pm}$ and $H$ are almost degenerate and they are lighter than $A$.
This implies that all of the additional Higgs bosons are heavier than $580$ GeV in Type-II, and it will be turned out that these are enough heavy to realize the alignment in our scenario.

We have also checked that the behaviors of $\lambda_{i}$ couplings are almost same for various $\tan{\beta}$ as long as $\tan{\beta}\lesssim 20$ and $\tan{\beta}$ is not close to unity.
In Type-II and Y, bottom Yukawa coupling is enhanced if $\tan{\beta}$ is large, and it may break the twisted custodial symmetry, however if the masses of the additional Higgs bosons are several hundred GeV, large $\tan{\beta}$ regions are excluded by $A\rightarrow \tau\tau$ and $A\rightarrow b\bar{b}$ decay modes \cite{Aaboud:2017sjh, Aad:2019zwb, Arbey:2017gmh}.
In Type-X, tau Yukawa coupling is enhanced when we take large $\tan{\beta}$, and such parameter regions are still allowed even when the masses of additional Higgs bosons are several hundred GeV.
However, the following discussions are also valid when $\tan{\beta}\lesssim 20$, and we do not discuss larger $\tan{\beta}$ scenario in Type-X in this paper.
We will discuss about the case of $\tan{\beta}\simeq 1$ later.

\begin{figure}[t]
\centering
\includegraphics[width=150mm]{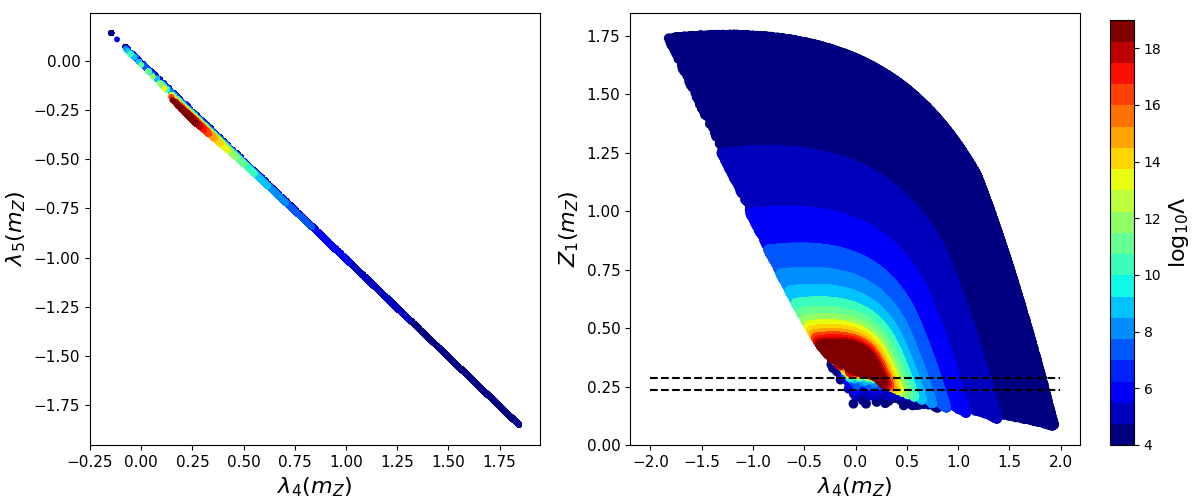}
\caption{(Left) Values of $\lambda_{4}$ and $\lambda_{5}$ at the electroweak scale for $\tan{\beta}=5$ in Type-I. (Right) Possible values of $Z_{1}$ and $\lambda_{4}$ at the electroweak scale. The points in different colors correspond to the different value of $\Lambda$ [GeV]. Dashed line shows the value of $Z_{1}$ which corresponds to $m_{h}=125\pm 6$ GeV in the alignment limit.}
\label{fig: tc_lam45}
\end{figure}

The left panel of Fig.~\ref{fig: tc_lam45} shows the possible values of $\lambda_{4}(m_{Z})$ and $\lambda_{5}(m_{Z})$. 
We can see that this scenario can be valid up to $\Lambda=10^{19}$ GeV, and a part of the twisted-custodial conditions in Eq.~\eqref{cBp}, $\lambda_{4}(m_{Z})=-\lambda_{5}(m_{Z})$, is approximately valid independently of the value of $\Lambda$.
The sign of $\lambda_{4}(m_{Z})$ tends to be positive, and its value converges to a small region if we take $\Lambda$ to close to the Planck scale.

The stability of $\lambda_{4}+\lambda_{5}=0$ can be understood by the form of the beta function,
\begin{align}
\text{Type-I:}\ 16\pi^{2}\frac{d(\lambda_{4}+\lambda_{5})}{d\ln{\mu}}
&=
2(\lambda_{1}+\lambda_{2}+4\lambda_{3}+2\lambda_{4}+4\lambda_{5})(\lambda_{4}+\lambda_{5})-3(3g^{2}+g'^{2})(\lambda_{4}+\lambda_{5}) \notag \\
&\quad
+2(3y_{t}^{2}+3y_{b}^{2}+y_{\tau}^{2})(\lambda_{4}+\lambda_{5})+3g^{2}g'^{2}.
\end{align}
As we have already mentioned, $g'$ and $y_{i}$ break the twisted custodial symmetry.
Even if $\lambda_{4}+\lambda_{5}=0$ at an initial point, $\lambda_{4}+\lambda_{5}$ is generated via the $g^{2}g'^{2}$ term.
However, this violating effect is negligible.
In Type-II and Y, we also have the $y_{t}^{2}y_{b}^{2}$ contribution to $\lambda_{4}+\lambda_{5}$. However, we confirmed that this effect is also small, and $\lambda_{4}+\lambda_{5}= 0$ is approximately valid at the electroweak scale.

The right panel of Fig.~\ref{fig: tc_lam45} shows the predicted value of $Z_{1}(m_{Z})$.
As we have discussed, $\lambda_{4}(m_{Z})$ tends to be positive and its value looks to converge to a small region.
These behavior can be understood by looking at the predicted value of $Z_{1}(m_{Z})$.
In the alignment limit, the mass of $h(125)$ is given by
\begin{align}
m_{h}^{2} = Z_{1}v^{2} &= [\lambda_{1}\cos^{4}{\beta}+\lambda_{2}\sin^{4}{\beta}+2\lambda_{345}\sin^{2}{\beta}\cos^{2}{\beta}]v^{2} \notag \\
&\simeq [\lambda_{1}\cos^{4}{\beta}+\lambda_{2}\sin^{4}{\beta}+2\lambda_{3}\sin^{2}{\beta}\cos^{2}{\beta}]v^{2},
\end{align}
where we have used $\lambda_{4}+\lambda_{5}\simeq0$ in the last equality.
To reproduce $m_{h}\simeq 125$ GeV, $Z_{1}$ should be $m_{h}^{2}/v^{2}\simeq 0.26$ in the alignment regions.
However, if $\lambda_{4}$ is negative, the vacuum stability condition;
\begin{align}
\sqrt{\lambda_{1}\lambda_{2}}+\lambda_{3}+\lambda_{4}-\lambda_{5}
\simeq\sqrt{\lambda_{1}\lambda_{2}}+\lambda_{3}+2\lambda_{4}>0,
\end{align}
sets the minimum value of $Z_{1}$ as
\begin{align}
Z_{1} &\simeq (\sqrt{\lambda_{1}}\cos^{2}{\beta}-\sqrt{\lambda_{2}}\sin^{2}{\beta})^{2}+2(\sqrt{\lambda_{1}\lambda_{2}}+\lambda_{3})\sin^{2}{\beta}\cos^{2}{\beta} \notag \\
&> (\sqrt{\lambda_{1}}\cos^{2}{\beta}-\sqrt{\lambda_{2}}\sin^{2}{\beta})^{2}-4\lambda_{4}\sin^{2}{\beta}\cos^{2}{\beta}. \label{Z1_bound}
\end{align}
This condition excludes $Z_{1}(m_{Z})\simeq 0.26$ for almost all values of negative $\lambda_{4}(m_{Z})$ in the alignment region.
Even positive $\lambda_{4}$, possible parameters with $Z_{1}\simeq 0.26$ are limited when $\Lambda$ is very high scale.
This is why the value of $\lambda_{4}(m_{Z})$ converges to a small region with $m_{h}\simeq 125$ GeV in the approximately alignment.

\begin{figure}[t]
\centering
\includegraphics[width=150mm]{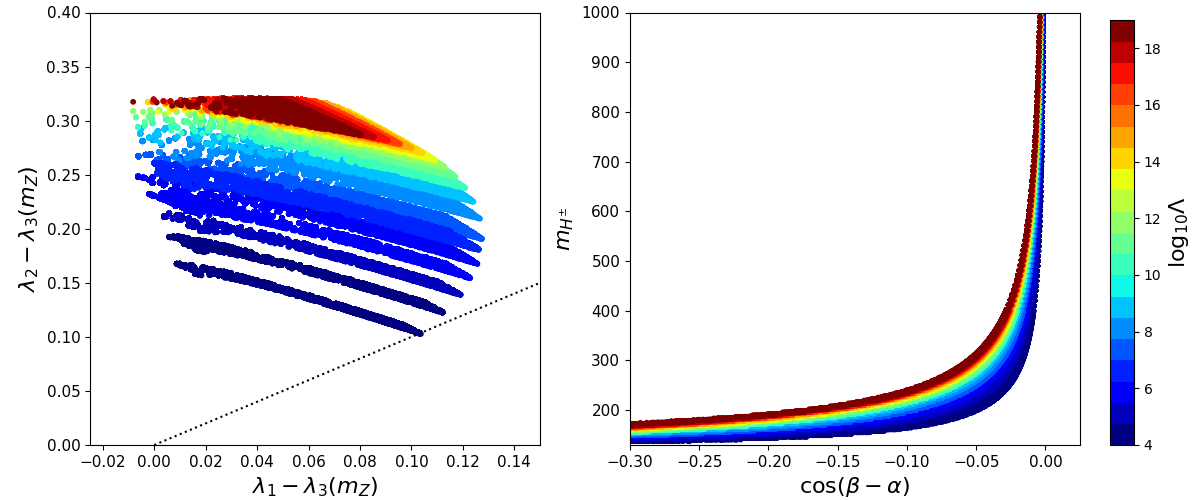}
\caption{(Left) Values of $\lambda_{1}-\lambda_{3}$ and $\lambda_{2}-\lambda_{3}$ at the electroweak scale for $\tan{\beta}=5$ in Type-I. The dashed line shows the parameter points where $\lambda_{1}=\lambda_{2}$. (Right) The decoupling behavior in the high-scale twisted-custodial-symmetric scenario. The points in different colors correspond to the different value of $\Lambda$ [GeV].}
\label{fig: tc_lamdif}
\end{figure}

The left panel of Fig.~\ref{fig: tc_lamdif} shows the possible values of $\lambda_{1}-\lambda_{3}$ and $\lambda_{2}-\lambda_{3}$ at the electroweak scale.
As we can see, both $\lambda_{1}-\lambda_{3}$ and $\lambda_{2}-\lambda_{3}$ take non-zero values, so that $\lambda_{1}\neq\lambda_{3}$ and $\lambda_{2}\neq\lambda_{3}$.
Furthermore, most of the parameter points are away from the dotted line which indicates $\lambda_{1}=\lambda_{2}$, so that the second condition $\lambda_{1}=\lambda_{2}=\lambda_{3}$ is violated at the electroweak scale.
This violation generates the off-diagonal component of the mass matrix for the CP-even scalars, and it predicts deviations in the couplings of $h(125)$ with various SM particles from those in the SM. 

The right panel of Fig.~\ref{fig: tc_lamdif} shows the decoupling behaviors in this scenario. 
We can see that $m_{H^{\pm}}\gtrsim 300$ GeV is enough heavy to achieve the alignment which satisfy the current experimental data \cite{ATLAS:2019slw, Sirunyan:2018koj}.
Although $Z_{6}$ is generated via RG running, its value is not so large comparing with the possible values which are allowed at the electroweak scale under the theoretical constraints.
Therefore, the decoupling-like behaviors appear even with relatively light additional Higgs bosons.
Thus, the alignment can be approximately realized without decoupling of the additional Higgs bosons due to the twisted custodial symmetry of the Higgs potential at the scale $\Lambda$.
In the Type-II scenario, same argument is valid, however $m_{H^{\pm}}\lesssim 580$ GeV is excluded by the constraint of $b\rightarrow s\gamma$ \cite{Misiak:2017bgg}.

\section{Predictions from the boundary conditions}
\label{sec:predictions}
\subsection{Mass spectrum of the additional Higgs bosons}
In this subsection, we analyze the prediction on the mass spectrum of the additional Higgs bosons in the twisted-custodial symmetric scenario at the high energy scale $\Lambda$.
In the alignment region, $\sin{(\beta-\alpha)}\simeq 1$, the masses of the additional Higgs bosons are given in terms of the parameters in the $\mathbb{Z}_{2}$ basis,
\begin{align}
m_{H}^{2} &\simeq M^{2}+\frac{1}{4}(\lambda_{1}+\lambda_{2}-2\lambda_{345})v^{2}\sin^{2}{2\beta}, \\
m_{H^{\pm}}^{2} &= M^{2}-\frac{1}{2}(\lambda_{4}+\lambda_{5})v^{2}, \\
m_{A}^{2} &= M^{2}-\lambda_{5}v^{2}.
\end{align}

\begin{figure}[t]
\centering
\includegraphics[width=150mm]{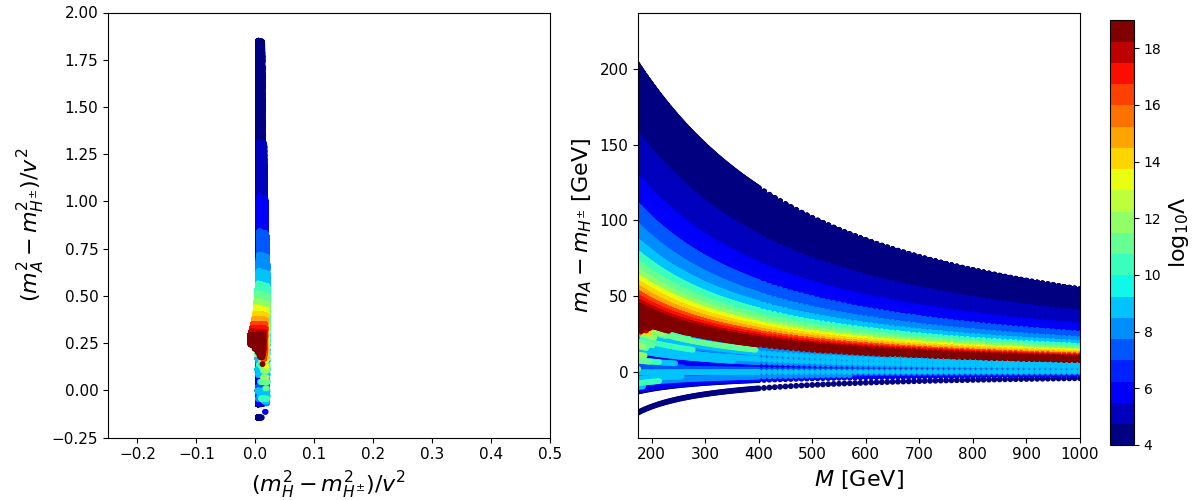}
\caption{Possible mass differences among the additional Higgs bosons for $\tan{\beta}=5$ in Type-I. The left figure shows $M$-independent mass-squared differences. The right figure shows the mass difference between $A$ and $H^{\pm}$ as a function of $M$. The points in different colors correspond to the different value of $\Lambda$ [GeV].}
\label{fig: tc_massdiff}
\end{figure}

The left panel of Fig.~\ref{fig: tc_massdiff} shows the predicted mass squared differences in this scenario.
We can see that $A$ tends to be heavier than $H^{\pm}$, while $H$ is approximately degenerate with $H^{\pm}$ in the almost parameter spaces.
%
To understand the prediction of the mass spectrum, we simplify the mass difference of the additional Higgs bosons using the condition $\lambda_{4}+\lambda_{5}\simeq 0$ which is valid even at the electroweak scale as shown in the left panel of Fig.~\ref{fig: tc_lam45},
\begin{align}
\frac{m_{A}^{2}-m_{H^{\pm}}^{2}}{v^{2}} = \lambda_{4}, \quad
\frac{m_{H}^{2}-m_{H^{\pm}}^{2}}{v^{2}} = (\lambda_{1}+\lambda_{2}-2\lambda_{3})\cot^{2}{\beta}\left(\frac{1}{1+\cot^{2}{\beta}}\right)^{2}. \label{mass_sq_dif}
\end{align}
The positivity and convergence behavior of the squared mass difference between $A$ and $H^{\pm}$ are the consequences of the prediction on $\lambda_{4}(m_{Z})$ which has been discussed in the previous section.
Since $\lambda_{1}-\lambda_{3}$ and $\lambda_{2}-\lambda_{3}$ take non-zero values as shown in Fig.~\ref{fig: tc_lamdif}, the mass squared difference between $H$ and $H^{\pm}$ is not zero.
However, this difference becomes small if $\tan{\beta}$ is not close to unity because $\lambda_{1}-\lambda_{3}$ and $\lambda_{2}-\lambda_{3}$ are ${\cal O}(10^{-1})$ and $\cot^{2}{\beta}$ suppresses the possible mass squared difference. 
Therefore, the following mass spectrum is predicted in this scenario,
\begin{align}
m_{A} \gtrsim m_{H}\simeq m_{H^{\pm}}.
\end{align}
Then, we can see that the condition (B) given in Eq~\eqref{condi_B} is realized, and the smallness of $\Delta T$ can be explained as a consequence of the twisted custodial symmetry at the scale $\Lambda$.

The right panel of Fig.~\ref{fig: tc_massdiff} shows the behavior of the mass difference $\Delta m=m_{A}-m_{H^{\pm}}$, where $\Delta m$ can be written as
\begin{align}
\Delta m \simeq M\left[\sqrt{1+\frac{\lambda_{4}v^{2}}{M^{2}}}-1\right], \quad (\mu=m_{Z}). \label{delta_m}
\end{align}
If we take the decoupling limit $M\rightarrow \infty$, $\Delta m$ is close to zero as depicted in the right panel of Fig.~\ref{fig: tc_massdiff}.
We note that the mass squared differences in Eq.~\eqref{mass_sq_dif} are $M$ independent quantities, while $\Delta m$ is $M$ dependent quantity.
If we determine $\Delta m$ and masses of $H$ and $H^{\pm}$, we can determine $\lambda_{4}$ and it imposes the upper bound of $\Lambda$.

We mention here the case of $\tan{\beta}\simeq 1$.
In this case, there is no $\cot^{2}{\beta}$ suppression in Eq.~\eqref{mass_sq_dif}, and $m_{H}-m_{H^{\pm}}$ can be taken about $50$ GeV as the maximum value.
However, these low $\tan{\beta}$ regions are constrained by the $b\rightarrow s\gamma$ experiments, and $m_{H^{\pm}} \gtrsim 600 {\rm GeV}$ even in the Type-I and X \cite{Arbey:2017gmh}.

\subsection{Deviations in the Higgs boson couplings}
In this subsection, we discuss deviations in the SM-like Higgs boson couplings with gauge bosons, quarks and leptons \cite{Kanemura:2014bqa} in our scenario.
It is convenient to define the scaling factors by normalizing the coupling constants of the SM Higgs boson,
\begin{align}
{\cal L}_{int}
=
\kappa_{V}h\left(\frac{m_{W}^{2}}{v}W^{+\mu}W_{\mu}^{-}+\frac{m_{Z}^{2}}{2v}Z^{\mu}Z_{\mu}\right)
-\sum_{f=u,d,e}\kappa_{f}h\frac{m_{f}}{v}\overline{f}f. \label{L_int}
\end{align}
From Eqs~\eqref{int_gauge} and \eqref{int_fermion}, $\kappa$ factors are given at tree level by
\begin{align}
\kappa_{V}=\sin{(\beta-\alpha)}, \quad \kappa_{f}=\xi_{h}^{f}=\sin{(\beta-\alpha)}+\xi_{f}\cos{(\beta-\alpha)}.
\end{align}
When $\kappa_{V}=\kappa_{f}=1$, the couplings of the SM-like Higgs boson take the SM values, and this SM-like limit can be achieved when $\sin{(\beta-\alpha)}=1$.
As we have seen in Sec.~\ref{sec:conditions}, the alignment is approximately realized at the electroweak scale, and this implies that $\kappa_{V}$ is necessarily close to the SM value in our scenario.
For example, $|\cos{(\beta-\alpha)}|\le 0.06$ when $\tan{\beta}=5$ and $m_{H^{\pm}}\ge 300$ GeV in Type-I as in Fig.~\ref{fig: tc_lamdif}, and it corresponds to $\kappa_{V}\ge 0.998$. It is difficult to measure this ${\cal O}(10^{-1})\%$ deviation of $\kappa_{V}$ even in the future precision measurement such as the HL-LHC and the ILC.
However, $\kappa_{f}$ can be more deviated from unity, because deviations are enhanced by $\tan{\beta}$ except for Type-I. Furthermore, the directions of modifications for $\kappa_{d}$ and  $\kappa_{e}$ are highly different in four types of Yukawa interactions.
Therefore, we can discriminate the type of 2HDMs through the precise measurement of $\kappa_{f}$ \cite{Kanemura:2014bqa}.

\begin{figure}[t]
\centering
\includegraphics[width=150mm]{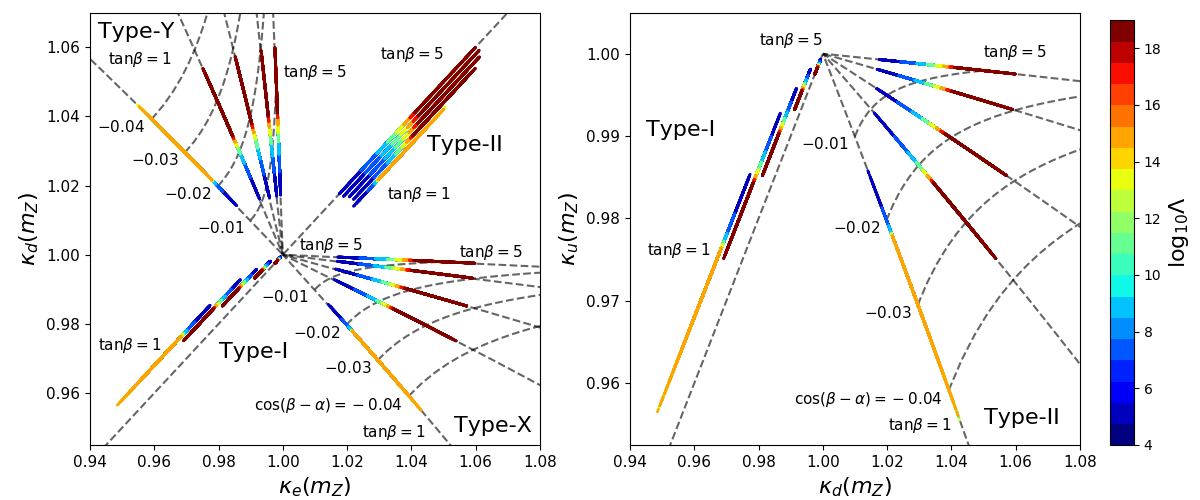}
\caption{Predictions on the scaling factors of the SM-like Higgs boson in 2HDMs. The left figure shows the predicted values of $\kappa_{d}$ and $\kappa_{e}$. The right figure shows those of $\kappa_{u}$ and $\kappa_{d}$. $\tan{\beta}=1, 1.5, 2, 3$ and $5$ with $600 {\rm GeV} \le M \le 700 {\rm GeV}$ are chosen as  representatives. The points in different colors correspond to the different value of $\Lambda$ [GeV].}
\label{fig: tc_kappa}
\end{figure}

The left panel of Fig.~\ref{fig: tc_kappa} shows the predicted values of $\kappa_{e}$ and $\kappa_{d}$ for the each type of 2HDMs.
We have plotted the predicted points for $\tan{\beta}=1,1.5, 2, 3$ and $5$ with $600\ {\rm GeV} \le M \le 700$ GeV.
For illustration purpose, we slightly shift lines along with $\kappa_{x}=\kappa_{y}$ in order to show $\tan{\beta}$ dependence in Type-I and II.
From the definition of the scaling factors in Eq.~\eqref{L_int}, the point $\kappa_{d}=\kappa_{e}=1$ corresponds to the SM-like limit.
As we can see, $\Lambda$ can be taken to the Planck scale if $\tan{\beta}\ge 1.5$. When $\tan{\beta}=1$, we cannot take $\Lambda$ to be the Planck scale and its maximal value is around $10^{14}$ GeV.
This is because the square of the top Yukawa coupling becomes larger than $4\pi$ during RG evolution when $\tan{\beta}=1$, and there is no solution above $\mu=10^{14}$ GeV independently of the values of $\lambda_{i}(\Lambda)$.
In Type-II, both of $\kappa_{d}$ and $\kappa_{e}$ deviate from SM values because
$\xi_{d}=\xi_{e}=-\tan{\beta}$ enhance these deviations.
In Type-X (Y), the modification of $\kappa_{e}\ (\kappa_{d})$ is enhanced, while $\kappa_{d}\ (\kappa_{e})$ closes to unity when we take larger value of $\tan{\beta}$ because its deviation proportional to $\cot{\beta}$.
In Type-I, both $\kappa_{d}$ and $\kappa_{e}$ approach to unity when we take larger value of $\tan{\beta}$.

The right panel of Fig.~\ref{fig: tc_kappa} shows the predicted values of $\kappa_{u}$ and $\kappa_{d}$.
We slightly shift lines along with $\kappa_{x}=\kappa_{y}$ in order to show $\tan{\beta}$ dependence in Type-I as in the left panel. The point $\kappa_{u}=\kappa_{d}=1$ corresponds to the SM-like limit.
Although the predicted values of $\cos{(\beta-\alpha)}$ depend on the type of 2HDM, we only show the results in Type-I and II, because the scaling factors $\kappa_{u}$ and $\kappa_{d}$ are same in Type-I (II) and X (Y), and the difference of predicted values between Type-I (II) and X (Y) which comes from the tau Yukawa coupling in the beta functions is almost negligible.
The modification of $\kappa_{u}$ is proportional to $\cot{\beta}$ independently of the types of 2HDM, and it closes to unity when we take larger value of $\tan{\beta}$.
In Type-II and Y, the modification of $\kappa_{d}$ is enhanced through $\tan{\beta}$, while $\kappa_{d}$ also approachs to unity by the $\cot{\beta}$ suppression in Type-I and X.

The possible deviations of $\kappa_{f}$ are determined from the value of $\cos{(\beta-\alpha)}$ for fixed $\tan{\beta}$, and it is generated from the violating effect of the twisted custodial symmetry under the RG evolution.
Therefore, modifications of the scaling factors become large when we take $\Lambda$ to higher scale, and we can expect about $5\%$ deviations of $\kappa_{d}$ and $\kappa_{e}$ when $\Lambda$ is the Planck scale.
We note that this several percent of deviations in the scaling factors can be tested at the future HL-LHC and ILC experiments, and we can investigate the possible scale $\Lambda$ through the precise measurement of $\kappa_{f}$.

Finally, we would like to mention about the difference of our results and previous works.
Discriminations of extended Higgs models through the precise measurement of the SM-like Higgs boson couplings have been studied at tree level \cite{Kanemura:2014bqa} and the one-loop level \cite{Kanemura:2018yai, Kanemura:2019kjg}.
In these works, the masses of the additional Higgs bosons, $M$ and the mixing angles are considered to be free parameters, and they are scanned under the theoretical and experimental constraints.
However, in our scenario, these values are predicted by the values of $\lambda_{i}$ at $\Lambda$, and the modifications of $\kappa_{f}$ are related to the possible scale $\Lambda$.
Therefore, we can utilize the precision measurement of the SM-like Higgs boson couplings not only to discriminate the types of the Yukawa interactions  but also to investigate the new physics scale $\Lambda$, where the global symmetry in the Higgs potential is restored and a fundamental theory should appear.

\section{Discussion and conclusions}
\label{sec:summary}
One of the signatures of this scenario is the mass spectrum of the additional Higgs bosons.
Especially, the mass difference among CP-odd and lighter states is important observable which determines the upper bound of $\Lambda$.
It can be separately measured by directly discovering $H^{\pm}$ and $A$ through the decay processes such as $H^{\pm}\rightarrow tb$ \cite{Aaboud:2018cwk}, $A\rightarrow \tau\tau$ \cite{Aaboud:2017sjh, Cepeda:2019klc} and so on.
We note that, if the scale $\Lambda$ is not so high and $\Delta m$ takes ${\cal O}(10^{2})$ GeV, we can determine $\Delta m$ using the same-sign pair production process of singly charged Higgs bosons ($pp\rightarrow W^{\pm*}W^{\pm*}jj\rightarrow H^{\pm}H^{\pm}jj$) whose cross section is proportional to the squared mass difference $m_{A}^{2}-m_{H^{\pm}}^{2}$ in the alignment limit \cite{Aiko:2019mww, Arhrib:2019ywg}.
For the small mass difference, it was pointed out that the non-decoupling effects in the $H^{\pm}W^{\mp}Z$ vertex are useful to study this mass difference \cite{Kanemura:1997ej, Kanemura:1999tg}.

This scenario predicts the approximately aligned Higgs boson couplings with gauge bosons which satisfy the current LHC data.
However, we can still investigate the predicted small mixing between the CP-even states through the precise measurement of the couplings of the SM-like Higgs boson and fermions.
When $\Lambda$ is close to the Planck scale, several percent of deviations of $\kappa_{f}$ are predicted, and such deviations can be tested at the future HL-LHC and ILC experiments.

An interesting application of our scenario would be electroweak baryogenesis. In our scenario, alignment $\sin{(\beta-\alpha)}\simeq 1$ is naturally realized with relatively light additional Higgs bosons.
Such a non-decoupling situation causes strongly first order phase transition, which is required for successful electroweak baryogenesis.
In our scenario, CP violation in the Higgs potential should be small at the electroweak scale due to the twisted custodial symmetry at $\Lambda$.
Thus, the origin of the CP violation should be in the Yukawa interactions relaxing the constraint from the softly-broken discrete symmetry.
Along this line, our scenario can be extended to a viable scenario for electroweak baryogenesis.

In this paper, we have utilized the one-loop beta functions to study the predictions on the electroweak scale observables because the values of Yukawa couplings at the scale $\Lambda$ can be determined independently of $\lambda_{i}$.
At the two-loop level, beta functions of Yukawa couplings depend on $\lambda_{i}$ and we need to make additional parameter scan to reproduce the correct fermion masses at the electroweak scale.
We leave this improvement for future work.

We have investigated a new scenario of 2HDMs where the current experimental data for the electroweak rho parameter and those for the Higgs boson couplings can be explained as a consequence of the global symmetry of the Higgs potential at the high energy scale $\Lambda$.
We have assumed that the twisted custodial symmetry results from some unknown theory at the scale $\Lambda$ and analyzed the violating effects of the rho parameter and the SM-like Higgs boson couplings.
We found that this scenario can be valid up to $\Lambda=10^{19}$ GeV and both small $\Delta T$ parameter and aligned Higgs boson couplings can be explained even when the masses of the additional Higgs bosons are around the electroweak scale.
This scenario predicts characteristic mass spectrum of the additional Higgs bosons, where the CP-odd Higgs boson is heavier than other Higgs bosons.
Furthermore, the mass squared difference between $A$ and $H^{\pm}$ converges to a definite value if $\Lambda$ is as large as the Planck scale.
In this scenario, alignment is approximately realized at the electroweak scale, and the Higgs-gauge couplings are close to the SM values.
However the modifications of the couplings between the SM-like Higgs boson and fermions are sensitive to the violation of alignment, and several percent of deviations are predicted when $\Lambda$ is close to the Planck scale.

\acknowledgments
M. A. was supported in part by the Sasakawa Scientific Research Grant from The Japan Science Society. S. K. was supported in part by JSPS, Grant-in-Aid for Scientific Research, 18F18321, 16H06492, 18H04587, 18F18022 and 20H00160.
\clearpage

\appendix
\section{Parameters of the Higgs potential in the Higgs basis}
\label{Apendix: Higgs-basis-params}
We here list the relations between the parameters of the Higgs potential in the Higgs basis and those in the $\mathbb{Z}_{2}$ basis in the softly-broken $\mathbb{Z}_{2}$ symmetric scenario \cite{Gunion:2005ja, Davidson:2005cw, Haber:2015pua, Bernon:2015qea, Boto:2020wyf};
\begin{align}
Y_{1}^{2} &= m_{11}^{2}\cos^{2}{\beta} + m_{22}^{2}\sin^{2}{\beta} - m_{12}^{2}\sin{2\beta}, \\
Y_{2}^{2} &= m_{11}^{2}\sin^{2}{\beta} + m_{22}^{2}\cos^{2}{\beta} + m_{12}^{2}\sin{2\beta}, \\
Y_{3}^{2} &= \frac{1}{2}(m_{11}^{2} - m_{2}^{22})\sin{2\beta} + m_{12}^{2}\cos{2\beta}, \\
Z_{1} &= \lambda_{1}\cos^{4}{\beta} + \lambda_{2}\sin^{4}{\beta} + \frac{1}{2}\lambda_{345}\sin^{2}{2\beta}, \\
Z_{2} &= \lambda_{1}\sin^{4}{\beta} + \lambda_{2}\cos^{4}{\beta} + \frac{1}{2}\lambda_{345}\sin^{2}{2\beta}, \\
Z_{3} &= \frac{1}{4}\sin^{2}{2\beta}[\lambda_{1} + \lambda_{2} -2\lambda_{345}] + \lambda_{3}, \\
Z_{4} &= \frac{1}{4}\sin^{2}{2\beta}[\lambda_{1} + \lambda_{2} -2\lambda_{345}] + \lambda_{4}, \\
Z_{5} &= \frac{1}{4}\sin^{2}{2\beta}[\lambda_{1} + \lambda_{2} -2\lambda_{345}] + \lambda_{5}, \\
Z_{6} &= -\frac{1}{2}\sin{2\beta}[\lambda_{1}\cos^{2}{\beta} - \lambda_{2}\sin^{2}{\beta} - \lambda_{345}\cos{2\beta}], \\
Z_{7} &= -\frac{1}{2}\sin{2\beta}[\lambda_{1}\sin^{2}{\beta} - \lambda_{2}\cos^{2}{\beta} + \lambda_{345}\cos{2\beta}],
\end{align}
and inversely
\begin{align}
m_{11}^{2} &= Y_{1}^{2}\cos^{2}{\beta} + Y_{2}^{2}\sin^{2}{\beta} + Y_{3}^{2}\sin{2\beta}, \\
m_{22}^{2} &= Y_{1}^{2}\sin^{2}{\beta} + Y_{2}^{2}\cos^{2}{\beta} - Y_{3}^{2}\sin{2\beta}, \\
m_{12}^{2} &= -\frac{1}{2}(Y_{1}^{2} - Y_{2}^{2})\sin{2\beta} + Y_{3}^{2}\cos{2\beta}, \\
\lambda_{1} &= Z_{1}\cos^{4}{\beta} + Z_{2}\sin^{4}{\beta} + \frac{1}{2}Z_{345}\sin^{2}{2\beta}
- 2Z_{6}\sin{2\beta}\cos^{2}{\beta} - 2Z_{7}\sin{2\beta}\sin^{2}{\beta}, \\
\lambda_{2} &= Z_{1}\sin^{4}{\beta} + Z_{2}\cos^{4}{\beta} + \frac{1}{2}Z_{345}\sin^{2}{2\beta}
+ 2Z_{6}\sin{2\beta}\sin^{2}{\beta} + 2Z_{7}\sin{2\beta}\cos^{2}{\beta}, \\
\lambda_{3} &= \frac{1}{4}\sin^{2}{2\beta}[Z_{1} + Z_{2} -2Z_{345}] + Z_{3} + (Z_{6}-Z_{7})\sin{2\beta}\cos{2\beta}, \\
\lambda_{4} &= \frac{1}{4}\sin^{2}{2\beta}[Z_{1} + Z_{2} -2Z_{345}] + Z_{4} + (Z_{6}-Z_{7})\sin{2\beta}\cos{2\beta}, \\
\lambda_{5} &= \frac{1}{4}\sin^{2}{2\beta}[Z_{1} + Z_{2} -2Z_{345}] + Z_{5} + (Z_{6}-Z_{7})\sin{2\beta}\cos{2\beta}.
\end{align}
In the softly-broken $\mathbb{Z}_{2}$ symmetric scenario, we do not have $\lambda_{6}$ and $\lambda_{7}$, therefore not all of $Z_{i}$ are independent and they satisfy following relations,
\begin{align}
\lambda_{6} &= \frac{1}{2}\sin{2\beta}\left[Z_{1}\cos^{2}{\beta} - Z_{2} \sin^{2}{\beta} - Z_{345}\cos{2\beta}\right] + Z_{6}\cos{\beta}\cos{3\beta} + Z_{7}\sin{\beta}\sin{3\beta}=0, \\
\lambda_{7} &= \frac{1}{2}\sin{2\beta}\left[Z_{1}\sin^{2}{\beta} - Z_{2} \cos^{2}{\beta} + Z_{345}\cos{2\beta}\right] + Z_{6}\sin{\beta}\sin{3\beta} + Z_{7}\cos{\beta}\cos{3\beta}=0.
\end{align}

When the Higgs potential respects the twisted custodial symmetry, we have
\begin{align}
Z_{4}+Z_{5}=0\quad {\rm and}\quad Z_{6}=Z_{7}=0, 
\end{align}
and 
\begin{align}
\lambda_{6}+\lambda_{7}&=\frac{1}{2}(Z_{1}-Z_{2})\sin{2\beta}, \\
\lambda_{6}-\lambda_{7}&=\frac{1}{2}(Z_{1}+Z_{2}-2Z_{3})\sin{2\beta}\cos{2\beta}.
\end{align}
If $\beta\neq 0,\pi/4$ or $\pi/2$, $Z_{1}=Z_{2}=Z_{3}$ and we have
\begin{align}
\lambda_{1}=\lambda_{2}=\lambda_{3},\quad \lambda_{4}+\lambda_{5}=0.
\end{align}
If $\beta=\pi/4$, $\lambda_{6}-\lambda_{7}=0$ is satisfied independently of $\lambda_{i}$ and we only have
\begin{align}
\lambda_{1}=\lambda_{2}=\lambda_{3}.
\end{align}
When $\beta = 0$ or $\pi/2$, the Higgs potential is not changed except for the sign of $\mathbb{Z}_{2}$ softly-broken term, and the twisted custodial symmetry implies
\begin{align}
\lambda_{4}+\lambda_{5}=0,\quad {\rm and}\quad m_{12}^{2}=0.
\end{align}
In this scenario, only one of the doublets $\Phi_{1}$ or $\Phi_{2}$ obtains the VEV, and this model corresponds to the inert doublet model \cite{Deshpande:1977rw, Barbieri:2006dq} if all fermions couple to the doublet which acquires the VEV.

\section{One-loop renormalization group equations for the 2HDMs}
\label{Apendix: RGE}
We here list the RG equations of dimensionless couplings up to one-loop level for the 2HDMs with softly-broken $\mathbb{Z}_{2}$ symmetry \cite{Cheng:1973nv, Komatsu:1981xh, Branco:2011iw, Das:2015mwa, Basler:2017nzu}.
The beta functions of $SU(3)_{c}, SU(2)_{L}$ and $U(1)_{Y}$ gauge couplings, $g_{s}, g$ and $g'$ are independent of types of Yukawa couplings and given by
\begin{align}
16\pi^{2} \beta_{g_{s}} &= \left(-11+\frac{4}{3}n_{g}\right)=-7g_{s}^{3}, \\
16\pi^{2} \beta_{g} &= \left(-\frac{22}{3}+\frac{4}{3}n_{g}+\frac{1}{6}n_{d}\right)=-3g^{3}, \\
16\pi^{2} \beta_{g'} &= \left(\frac{20}{9}n_{g}+\frac{1}{6}n_{d}\right)=7g'^{3},
\end{align}
where $n_{g}$ is the number of the generation of the fermions and $n_{d}$ is the number of the scalar doublets. In the 2HDMs, $n_{g}=3$ and $n_{d}=2$.

\subsection{Type-I model}
In the Type-I model, $\Phi_{1}$ does not couple to fermions while $\Phi_{2}$ does to all fermions.
The $\beta$ functions of $\lambda_{i}$ are given by
\begin{align}
16\pi^{2}\beta_{\lambda_{1}}
&=
12\lambda_{1}^{2}+4\lambda_{3}^{2}+4\lambda_{3}\lambda_{4}+2\lambda_{4}^{2}+2\lambda_{5}^{2}
+\frac{3}{4}(3g^{4}+g'^{4}+2g^{2}g'^{2})-3\lambda_{1}(3g^{2}+g'^{2}), \\
16\pi^{2}\beta_{\lambda_{2}}
&=
12\lambda_{2}^{2}+4\lambda_{3}^{2}+4\lambda_{3}\lambda_{4}+2\lambda_{4}^{2}+2\lambda_{5}^{2}
+\frac{3}{4}(3g^{4}+g'^{4}+2g^{2}g'^{2})-3\lambda_{2}(3g^{2}+g'^{2}) \notag \\
&\quad
+4(3y_{t}^{2}+3y_{b}^{2}+y_{\tau}^{2})\lambda_{2}-4(3y_{t}^{4}+3y_{b}^{4}+y_{\tau}^{4}), \\
16\pi^{2}\beta_{\lambda_{3}}
&=
(\lambda_{1}+\lambda_{2})(6\lambda_{3}+2\lambda_{4})+4\lambda_{3}^{2}+2\lambda_{4}^{2}+2\lambda_{5}^{2}
+\frac{3}{4}(3g^{4}+g'^{4}-2g^{2}g'^{2})-3\lambda_{3}(3g^{2}+g'^{2}) \notag \\
&\quad
+2(3y_{t}^{2}+3y_{b}^{2}+y_{\tau}^{2})\lambda_{3}, \\
16\pi^{2}\beta_{\lambda_{4}}
&=
2(\lambda_{1}+\lambda_{2})\lambda_{4}+8\lambda_{3}\lambda_{4}+4\lambda_{4}^{2}+8\lambda_{5}^{2}
+3g^{2}g'^{2}-3(3g^{2}+g'^{2})\lambda_{4} \notag \\
&\quad
+2(3y_{t}^{2}+3y_{b}^{2}+y_{\tau}^{2})\lambda_{4}, \\
16\pi^{2}\beta_{\lambda_{5}}
&=
2(\lambda_{1}+\lambda_{2}+4\lambda_{3}+6\lambda_{4})\lambda_{5}-3\lambda_{5}(3g^{2}+g'^{2})
+2(3y_{t}^{2}+3y_{b}^{2}+y_{\tau}^{2})\lambda_{5}.
\end{align}
The $\beta$ functions of Yukawa couplings are given by
\begin{align}
16\pi^{2}\beta_{y_{t}}
&=\left(-8g_{s}^{2}-\frac{9}{4}g^{2}-\frac{17}{12}g'^{2}+\frac{9}{2}y_{t}^{2}+\frac{3}{2}y_{b}^{2}+y_{\tau}^{2}\right)y_{t}, \\
16\pi^{2}\beta_{y_{b}}
&=\left(-8g_{s}^{2}-\frac{9}{4}g^{2}-\frac{5}{12}g'^{2}+\frac{3}{2}y_{t}^{2}+\frac{9}{2}y_{b}^{2}+y_{\tau}^{2}\right)y_{b}, \\
16\pi^{2}\beta_{y_{\tau}}
&=\left(-\frac{9}{4}g^{2}-\frac{15}{4}g'^{2}+3y_{t}^{2}+3y_{b}^{2}+\frac{5}{2}y_{\tau}^{2}\right)y_{\tau}.
\end{align}
\subsection{Type-II model}
In the Type-II model, $\Phi_{1}$ couples to down-type quarks and leptons while $\Phi_{2}$ does to up-type quarks.
The $\beta$ functions of $\lambda_{i}$ are given by
\begin{align}
16\pi^{2}\beta_{\lambda_{1}}
&=
12\lambda_{1}^{2}+4\lambda_{3}^{2}+4\lambda_{3}\lambda_{4}+2\lambda_{4}^{2}+2\lambda_{5}^{2}
+\frac{3}{4}(3g^{4}+g'^{4}+2g^{2}g'^{2})-3\lambda_{1}(3g^{2}+g'^{2}) \notag \\
&\quad
+4(3y_{b}^{2}+y_{\tau}^{2})\lambda_{1}-4(3y_{b}^{4}+y_{\tau}^{4}), \\
16\pi^{2}\beta_{\lambda_{2}}
&=
12\lambda_{2}^{2}+4\lambda_{3}^{2}+4\lambda_{3}\lambda_{4}+2\lambda_{4}^{2}+2\lambda_{5}^{2}
+\frac{3}{4}(3g^{4}+g'^{4}+2g^{2}g'^{2})-3\lambda_{2}(3g^{2}+g'^{2}) \notag \\
&\quad
+12y_{t}^{2}\lambda_{2}-12y_{t}^{4}, \\
16\pi^{2}\beta_{\lambda_{3}}
&=
(\lambda_{1}+\lambda_{2})(6\lambda_{3}+2\lambda_{4})+4\lambda_{3}^{2}+2\lambda_{4}^{2}+2\lambda_{5}^{2}
+\frac{3}{4}(3g^{4}+g'^{4}-2g^{2}g'^{2})-3\lambda_{3}(3g^{2}+g'^{2}) \notag \\
&\quad
+2(3y_{t}^{2}+3y_{b}^{2}+y_{\tau}^{2})\lambda_{3}-12y_{t}^{2}y_{b}^{2}, \\
16\pi^{2}\beta_{\lambda_{4}}
&=
2(\lambda_{1}+\lambda_{2})\lambda_{4}+8\lambda_{3}\lambda_{4}+4\lambda_{4}^{2}+8\lambda_{5}^{2}
+3g^{2}g'^{2}-3(3g^{2}+g'^{2})\lambda_{4} \notag \\
&\quad
+2(3y_{t}^{2}+3y_{b}^{2}+y_{\tau}^{2})\lambda_{4}+12y_{t}^{2}y_{b}^{2}, \\
16\pi^{2}\beta_{\lambda_{5}}
&=
2(\lambda_{1}+\lambda_{2}+4\lambda_{3}+6\lambda_{4})\lambda_{5}-3\lambda_{5}(3g^{2}+g'^{2})
+2(3y_{t}^{2}+3y_{b}^{2}+y_{\tau}^{2})\lambda_{5}.
\end{align}
The $\beta$ functions of Yukawa couplings are given by
\begin{align}
16\pi^{2}\beta_{y_{t}}
&=\left(-8g_{s}^{2}-\frac{9}{4}g^{2}-\frac{17}{12}g'^{2}+\frac{9}{2}y_{t}^{2}+\frac{1}{2}y_{b}^{2}\right)y_{t}, \\
16\pi^{2}\beta_{y_{b}}
&=\left(-8g_{s}^{2}-\frac{9}{4}g^{2}-\frac{5}{12}g'^{2}+\frac{1}{2}y_{t}^{2}+\frac{9}{2}y_{b}^{2}+y_{\tau}^{2}\right)y_{b}, \\
16\pi^{2}\beta_{y_{\tau}}
&=\left(-\frac{9}{4}g^{2}-\frac{15}{4}g'^{2}+3y_{b}^{2}+\frac{5}{2}y_{\tau}^{2}\right)y_{\tau}.
\end{align}
\subsection{Type-X model}
In the Type-X model $\Phi_{1}$ couples to leptons while $\Phi_{2}$ does to all quarks.
The $\beta$ functions of $\lambda_{i}$ are given by
\begin{align}
16\pi^{2}\beta_{\lambda_{1}}
&=
12\lambda_{1}^{2}+4\lambda_{3}^{2}+4\lambda_{3}\lambda_{4}+2\lambda_{4}^{2}+2\lambda_{5}^{2}
+\frac{3}{4}(3g^{4}+g'^{4}+2g^{2}g'^{2})-3\lambda_{1}(3g^{2}+g'^{2}) \notag \\
&\quad
+4y_{\tau}^{2}\lambda_{1}-4y_{\tau}^{4}, \\
16\pi^{2}\beta_{\lambda_{2}}
&=
12\lambda_{2}^{2}+4\lambda_{3}^{2}+4\lambda_{3}\lambda_{4}+2\lambda_{4}^{2}+2\lambda_{5}^{2}
+\frac{3}{4}(3g^{4}+g'^{4}+2g^{2}g'^{2})-3\lambda_{2}(3g^{2}+g'^{2}) \notag \\
&\quad
+4(3y_{t}^{2}+3y_{b}^{2})\lambda_{2}-4(3y_{t}^{4}+3y_{b}^{4}), \\
16\pi^{2}\beta_{\lambda_{3}}
&=
(\lambda_{1}+\lambda_{2})(6\lambda_{3}+2\lambda_{4})+4\lambda_{3}^{2}+2\lambda_{4}^{2}+2\lambda_{5}^{2}
+\frac{3}{4}(3g^{4}+g'^{4}-2g^{2}g'^{2})-3\lambda_{3}(3g^{2}+g'^{2}) \notag \\
&\quad
+2(3y_{t}^{2}+3y_{b}^{2}+y_{\tau}^{2})\lambda_{3}, \\
16\pi^{2}\beta_{\lambda_{4}}
&=
2(\lambda_{1}+\lambda_{2})\lambda_{4}+8\lambda_{3}\lambda_{4}+4\lambda_{4}^{2}+8\lambda_{5}^{2}
+3g^{2}g'^{2}-3(3g^{2}+g'^{2})\lambda_{4} \notag \\
&\quad
+2(3y_{t}^{2}+3y_{b}^{2}+y_{\tau}^{2})\lambda_{4}, \\
16\pi^{2}\beta_{\lambda_{5}}
&=
2(\lambda_{1}+\lambda_{2}+4\lambda_{3}+6\lambda_{4})\lambda_{5}-3\lambda_{5}(3g^{2}+g'^{2})
+2(3y_{t}^{2}+3y_{b}^{2}+y_{\tau}^{2})\lambda_{5}.
\end{align}
The $\beta$ functions of Yukawa couplings are given by
\begin{align}
16\pi^{2}\beta_{y_{t}}
&=\left(-8g_{s}^{2}-\frac{9}{4}g^{2}-\frac{17}{12}g'^{2}+\frac{9}{2}y_{t}^{2}+\frac{3}{2}y_{b}^{2}\right)y_{t}, \\
16\pi^{2}\beta_{y_{b}}
&=\left(-8g_{s}^{2}-\frac{9}{4}g^{2}-\frac{5}{12}g'^{2}+\frac{3}{2}y_{t}^{2}+\frac{9}{2}y_{b}^{2}\right)y_{b}, \\
16\pi^{2}\beta_{y_{\tau}}
&=\left(-\frac{9}{4}g^{2}-\frac{15}{4}g'^{2}+\frac{5}{2}y_{\tau}^{2}\right)y_{\tau}.
\end{align}
\subsection{Type-Y model}
In the Type-Y model, $\Phi_{1}$ couples to down-type quarks while $\Phi_{2}$ does to up-type quarks and leptons.
The $\beta$ functions of $\lambda_{i}$ are given by
\begin{align}
16\pi^{2}\beta_{\lambda_{1}}
&=
12\lambda_{1}^{2}+4\lambda_{3}^{2}+4\lambda_{3}\lambda_{4}+2\lambda_{4}^{2}+2\lambda_{5}^{2}
+\frac{3}{4}(3g^{4}+g'^{4}+2g^{2}g'^{2})-3\lambda_{1}(3g^{2}+g'^{2}) \notag \\
&\quad
+12y_{b}^{2}\lambda_{1}-12y_{b}^{4}, \\
16\pi^{2}\beta_{\lambda_{2}}
&=
12\lambda_{2}^{2}+4\lambda_{3}^{2}+4\lambda_{3}\lambda_{4}+2\lambda_{4}^{2}+2\lambda_{5}^{2}
+\frac{3}{4}(3g^{4}+g'^{4}+2g^{2}g'^{2})-3\lambda_{2}(3g^{2}+g'^{2}) \notag \\
&\quad
+4(3y_{t}^{2}+y_{\tau}^{2})\lambda_{2}-4(3y_{t}^{4}+y_{\tau}^{4}), \\
16\pi^{2}\beta_{\lambda_{3}}
&=
(\lambda_{1}+\lambda_{2})(6\lambda_{3}+2\lambda_{4})+4\lambda_{3}^{2}+2\lambda_{4}^{2}+2\lambda_{5}^{2}
+\frac{3}{4}(3g^{4}+g'^{4}-2g^{2}g'^{2})-3\lambda_{3}(3g^{2}+g'^{2}) \notag \\
&\quad
+2(3y_{t}^{2}+3y_{b}^{2}+y_{\tau}^{2})\lambda_{3}-12y_{t}^{2}y_{b}^{2}, \\
16\pi^{2}\beta_{\lambda_{4}}
&=
2(\lambda_{1}+\lambda_{2})\lambda_{4}+8\lambda_{3}\lambda_{4}+4\lambda_{4}^{2}+8\lambda_{5}^{2}
+3g^{2}g'^{2}-3(3g^{2}+g'^{2})\lambda_{4} \notag \\
&\quad
+2(3y_{t}^{2}+3y_{b}^{2}+y_{\tau}^{2})\lambda_{4}+12y_{t}^{2}y_{b}^{2}, \\
16\pi^{2}\beta_{\lambda_{5}}
&=
2(\lambda_{1}+\lambda_{2}+4\lambda_{3}+6\lambda_{4})\lambda_{5}-3\lambda_{5}(3g^{2}+g'^{2})
+2(3y_{t}^{2}+3y_{b}^{2}+y_{\tau}^{2})\lambda_{5}.
\end{align}
The $\beta$ functions of Yukawa couplings are given by
\begin{align}
16\pi^{2}\beta_{y_{t}}
&=\left(-8g_{s}^{2}-\frac{9}{4}g^{2}-\frac{17}{12}g'^{2}+\frac{9}{2}y_{t}^{2}+\frac{1}{2}y_{b}^{2}+y_{\tau}^{2}\right)y_{t}, \\
16\pi^{2}\beta_{y_{b}}
&=\left(-8g_{s}^{2}-\frac{9}{4}g^{2}-\frac{5}{12}g'^{2}+\frac{1}{2}y_{t}^{2}+\frac{9}{2}y_{b}^{2}\right)y_{b}, \\
16\pi^{2}\beta_{y_{\tau}}
&=\left(-\frac{9}{4}g^{2}-\frac{15}{4}g'^{2}+3y_{t}^{2}+\frac{5}{2}y_{\tau}^{2}\right)y_{\tau}.
\end{align}


\bibliography{bibRG}
\bibliographystyle{junsrt}

\end{document}